\documentclass[10pt, twocolumn, secnumarabic, amssymb, nobibnotes, footinbib, aps, pra, superscriptaddress]{revtex4-1}

\usepackage{graphicx}
\usepackage{dcolumn}
\usepackage{amsthm}
\usepackage{amsmath}
\usepackage{mathtools}
\usepackage{natbib}
\usepackage{color}
\usepackage{epstopdf}

\usepackage[english]{babel}

\begin{document}

\title{On dynamical stability of composite quantum particles: when entanglement is enough and when interaction is needed}

\author{Zakarya Lasmar}
\affiliation{Faculty of Physics, Adam Mickiewicz University, Umultowska 85, 61-614 Pozna\'n, Poland}

\author{Adam S. Sajna}
\affiliation{Faculty of Physics, Adam Mickiewicz University, Umultowska 85, 61-614 Pozna\'n, Poland}

\author{Su-Yong Lee}
\affiliation{School of Computational Sciences, Korea Institute for Advanced Study, Hoegi-ro 85, Dongdaemun-gu, Seoul 02455, Korea}

\author{Pawe\l{} Kurzy\'nski}   \email{pawel.kurzynski@amu.edu.pl}  
\affiliation{Faculty of Physics, Adam Mickiewicz University, Umultowska 85, 61-614 Pozna\'n, Poland}
\affiliation{Centre for Quantum Technologies, National University of Singapore, 3 Science Drive 2, 117543 Singapore, Singapore}

\date{\today}


\begin{abstract}
We consider an evolution of two elementary quantum particles and ask the question: under what conditions such a system behaves as a single object? It is obvious that if the attraction between the particles is stronger than any other force acting on them the whole system behaves as one. However, recent insight from the quantum information theory suggests that in bipartite systems it is not attraction per se that is responsible for the composite nature, but the entanglement between the parts. Since entanglement can be present between the subsystems that interacted in the past, but do not interact anymore, it is natural to ask when such an entangled pair behaves as a single object. We show that there are situations when entanglement is enough to observe single-particle behaviour. However, due to the no-signalling condition, in general an interaction, or a post-selective measurement, is necessary for a complex collective behaviour.    
\end{abstract}


\maketitle


\section{Introduction}

Composite particles naturally arise in systems of interacting elementary particles. However, recent studies on composite quantum particles suggest that the phenomenon of compositeness is not exactly due to the interaction, but rather due to the entanglement that is caused by the interaction \cite{Law}. Therefore, it is valid to ask whether non-interacting entangled subsystems can behave as a single stable quantum object. 

Studies on compositeness in quantum regime should take into account all the fundamental features of the theory, such as the wave-particle duality. An elementary quantum particle is also an elementary quantum wave. Whether the system manifests a particle or a wave nature depends on the choice of a physical property one wants to observe. Similarly, a composite quantum object should also exhibit the wave-particle duality. Therefore, the compositeness of quantum systems should be studied in two different types of experiments, the ones focusing on the particle-like behaviour and the others focusing on the wave-like one. 

The problem of compositeness of multipartite quantum systems is not new, but has been mostly studied in scenarios for which there are natural intra-particle interactions. However, our line of thought is rather related to the studies on compositeness within quantum optics \cite{PhotonicdeBroglieWaves95,Fonseca99,Sackett00,Fonseca01,Edamatsu02} and quantum information theory \cite{Law,e1,e2,e3,e4,e5,e6,e7,e8,e9,e10,e11,e12,e13,e14,e15,e16,e17,e18,e19,e20,e21,e22,e23}. This is because we focus on entanglement not on interaction. The field of quantum optics deals with photons that do not interact easily without a special mediator. On the other hand, the quantum information theory studies fundamental properties of entanglement. Nevertheless, in this work we also discuss effects known from solid state and cold atom physics.  

Although previous research on compositeness within quantum optics and quantum information theory proved that entanglement is necessary to observe various composite effects, it was not clearly determined when entanglement is a sufficient condition to observe the composite behaviour. In this work we show that the compositeness due to entanglement alone is conditioned on the property one wants to measure. We study three types of dynamics: free-evolution, interference in the Mach-Zehnder-like interferometer, and the Bloch oscillations on a lattice due to external linear potential. Moreover, we investigate how thermalization affects the stability of such a composite particle. The unconditional composite nature of entangled bipartite systems, i.e., the one that does not require a special type of measurements, is exhibited only in the first case. Finally, we observe that the origin of this effect stems from a variant of the no-signalling condition which states that interaction is needed whenever the dynamics of one subsystem depends on the behaviour of the other one.


\section{What is a particle?}

Let us start with a fundamental question that is relevant for us. {\it What is a particle?} A particle is a localized object. Localization in space is its fundamental property that distinguishes it from being a wave. An additional property of a particle is its velocity, or momentum, which describes how its localization changes in time. Apart from localization and velocity, a particle can possess additional properties (such as charge, spin, etc.), however in our discussion these additional properties do not play any significant role. We focus on localization and velocity. 

What does it mean that particle is localized? Practically speaking, localization implies that whenever one measures where the particle is, one finds that it can be found in an exact position in space, or, in case of a particle that has some volume, one finds that it can be found in some confined region. In other words, if one sets some number of particle detectors in different regions in space, a single particle will make only one of these detectors click. Therefore, a single particle can be associated with a single detector click. 

{\it What is a composite particle?} Here, we provide some basic definitions and assumptions. Firstly, since a single particle corresponds to a single click, the simplest definition of a composite particle is a system that is known to be made of more than one part, but still produces a single detector click if one asks about its position. This means that the constituents stay together (or close to each other, which results in a non-zero volume of the whole system).

In addition, it is intuitive to assume that a system composed of many subsystems, in principle describable by many parameters, behaves as a single particle if its spatial state can be effectively described by only a few parameters. Such situation naturally occurs in the presence of strong correlations. 
The stronger the correlations between the system's parameters, the easier it is to predict its behaviour.  
For example, if the constituents stay close to each other, the knowledge of the position of one element automatically gives some information about the positions of the remaining ones. Moreover, if the velocities of all the elements are the same, the initial relation between the positions will remain constant in time during a free motion.   

Finally, one can assume that the internal state of the composite particle does not change (although later we will drop this assumption). In this case one should aim to separate the evolution of the centre of mass from the other degrees of freedom. Then, one should freeze the evolution of the other degrees of freedom such that only the centre of mass evolves. 

\begin{figure}
\includegraphics[scale=0.30]{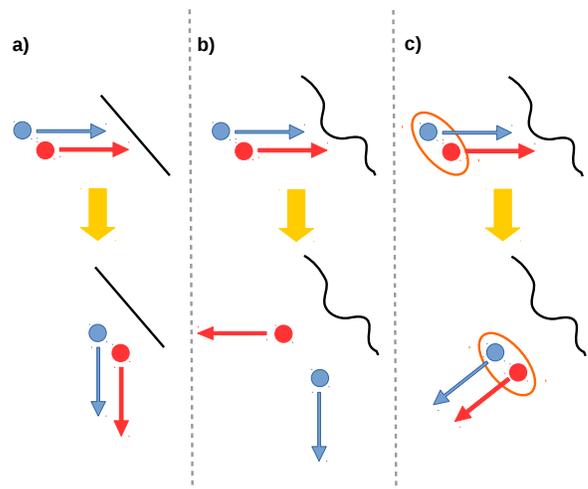}
\caption{Stability of a classical composite system (schematic representation). a) Two non-interacting particles with correlated momenta stay close to each other after scattering from a flat wall. The wall reflects both particles the same way, so the momenta stay correlated. b) Two non-interacting particles with correlated momenta drift away from each other after scattering from an irregular wall. Each particle is reflected at a different angle so the correlation between momenta is disturbed. c) Two interacting particles stay close to each other after scattering from an irregular wall. Although each particle is reflected at a different angle, the attraction keeps them together and guarantees that momenta stay correlated. \label{fig1}}
\end{figure}

Before we proceed, let us briefly discuss the problem of composite particles in classical physics. We consider a system made up of two elementary particles. As we noted above, the compositeness of this system depends on correlations between the constituents. Since the composite system must be localized in order to be called a particle (according to our definition), the elementary particles need to stay close to each other and their velocities need to be similar in order to keep the close distance between the positions. It is clear that for free evolution in which the velocities are constant the attractive intra-particle interaction is not necessary to keep the system localized. We only need specific correlations. This is also true in a case of non-free evolutions in which the composite system is subjected to external potentials that affect the particles in the same way, i.e., that cause the same change to their velocities. That way the spatial correlations between the particle positions are conserved and the initially localized composite particle stays localized in the future. However, the attractive intra-particle interaction is needed whenever the external potential affects the particles in a different way. In this case the change of velocities is different and if there is no additional force keeping the constituents together, the composite particle falls apart. The above situations are schematically depicted in Fig.~\ref{fig1}.


\section{Quantum particles}
 
Unlike classical particles, quantum particles exhibit the wave-particle duality. In case of a free evolution the duality means that an initially localized quantum particle will start to behave like a wave and will disperse over whole space. However, the measurement of its position will bring back the particle to a localized state, although in an indeterministic way. In other words, only a single random detector will click.
 
Next, let us consider composite quantum particles. We argued that in classical theory the compositeness of a bipartite system is related to the correlations between the positions and between the velocities of both elementary particles. However, in quantum theory one can either know the position or the velocity of each particle, but not both at the same time. This leads to the following problem: if at time $t_0$ one places two quantum particles in the same place, one perfectly correlates their positions, but knows nothing about their velocities. Because of that at time $t_1 > t_0$ the particles will disperse and their position states will get delocalized in space. The particles will disperse independently, therefore at time $t_1$ the knowledge of the position of one particle says nothing about the position of the other one. Moreover, it is likely that at time $t_1$ the particles will be far away from each other. Therefore, according to our definition, the system cannot be considered a composite particle. 

It is of course possible to prepare two quantum particles in a state being a product of two Gaussian wave-functions with the same average position and the same average momentum. Such a system will mimic the classical composite particle from the previous section, but it will not exhibit the wave-particle duality in the sense described above. We are therefore looking for special quantum states of two particles that evolve from a localized state to a delocalized one, but which assure that the two particles stay close to each other, so that a subsequent position measurement will find both particles in one place. Therefore, such states would have to exhibit some peculiar type of correlations.  

Interestingly, although it is impossible to determine the position and the velocity of each particle at the same time, it is possible to know the correlations between their positions and between their momenta (and as a result between their velocities). This effect is known as entanglement \cite{Horo}. Therefore, it is natural to speculate that entanglement may play some role in the studies of composite quantum particles. Indeed, in a number of works \cite{Law,e1,e2,e3,e4,e5,e6,e7,e8,e9,e10,e11,e12,e13,e14,e15,e16,e17,e18,e19,e20,e21,e22,e23} it was argued that two elementary fermions, or two elementary bosons, may behave like a single composite boson if they are sufficiently entangled. In the following sections we develop this idea and ask if the dynamics of such an entangled pair can be interpreted as a behaviour of a single particle. We focus on composite systems made of two entangled one-dimensional spin-less particles.

\section{Free evolution of entangled particles}

\subsection{Double Gaussian state}

We are going to consider Gaussian wave packets because of their simple mathematical description and interesting physical properties. Nevertheless, the conclusions drawn from this study will apply to a much more general class of states. A standard single-particle Gaussian packet centred around $x=0$ with initial momentum centred around $p=0$ is given by $\psi(x,t=0)={\cal N}\text{exp}(-x^2/2\sigma^2)$, where $\cal{N}$ is a normalisation factor and $\sigma/\sqrt{2}$ is a standard deviation. For free evolution the standard deviation changes in time as
\begin{equation}\label{sigma}
\Delta x (t) = \frac{1}{\sqrt{2}}\sqrt{\sigma^2 + \frac{\hbar^2 t^2}{m^2 \sigma^2}},
\end{equation}
where $m$ is the mass of the particle. For $t = \frac{m}{\hbar}\sigma^2$ the initial standard deviation increases by the factor of $\sqrt{2}$ and for $t\gg \frac{m}{\hbar}\sigma^2$ we can approximate $\Delta x (t) \approx \frac{\hbar}{\sqrt{2}m\sigma} t$. Therefore, the greater the initial variance, the slower the wave packet spreads.

Next, we define the following double Gaussian wave function of two one-dimensional particles
\begin{equation}\label{thestate}
\psi(x_1,x_2,t=0) = {\cal N}e^{-\frac{(x_1-x_2)^2}{4\sigma^2}}e^{-\frac{(x_1+x_2)^2}{4\Sigma^2}},
\end{equation}
where this time $\sigma/\sqrt{2}$ and $\Sigma/\sqrt{2}$ correspond to standard deviations of the relative position $\frac{x_1 - x_2}{\sqrt{2}}$ and of the position of the centre of mass $\frac{x_1 + x_2}{\sqrt{2}}$, respectively. The entanglement between the particles can be measured by the purity of a single particle density matrix, which in this case is given by \cite{Law}
\begin{equation}\label{purity}
P=\text{Tr}\{\rho_1(x_1,x'_1)^2\}=\text{Tr}\{\rho_2(x_2,x'_2)^2\} = \frac{2\sigma\Sigma}{\sigma^2 + \Sigma^2}.
\end{equation}
The state is separable ($P=1$) if and only if $\sigma=\Sigma$. In any other case the state is entangled ($P<1$). In the limit of strong entanglement ($P\ll 1$) either $\sigma \ll \Sigma$ or $\sigma \gg \Sigma$. The correlation properties of the double Gaussian wave function have been already discussed in \cite{Law}. Here, we study its dynamics. 

\subsection{Free evolution}

Consider an evolution of the state (\ref{thestate}) generated by the free particle Hamiltonian $H_{free}=\frac{p_1^2 + p_2^2}{2m}$. For simplicity we assume that both particles have the same mass. Since $H_{free}=H_+ + H_-$, where $H_{\pm} = \frac{(p_1 \pm p_2)^2}{4m}$, each part of (\ref{thestate}) evolves independently like a standard single-partite free Gaussian wave-packet. Therefore, the standard deviations of the centre of mass and of the relative position evolve analogously to (\ref{sigma}). 

\begin{figure}
\includegraphics[scale=0.32]{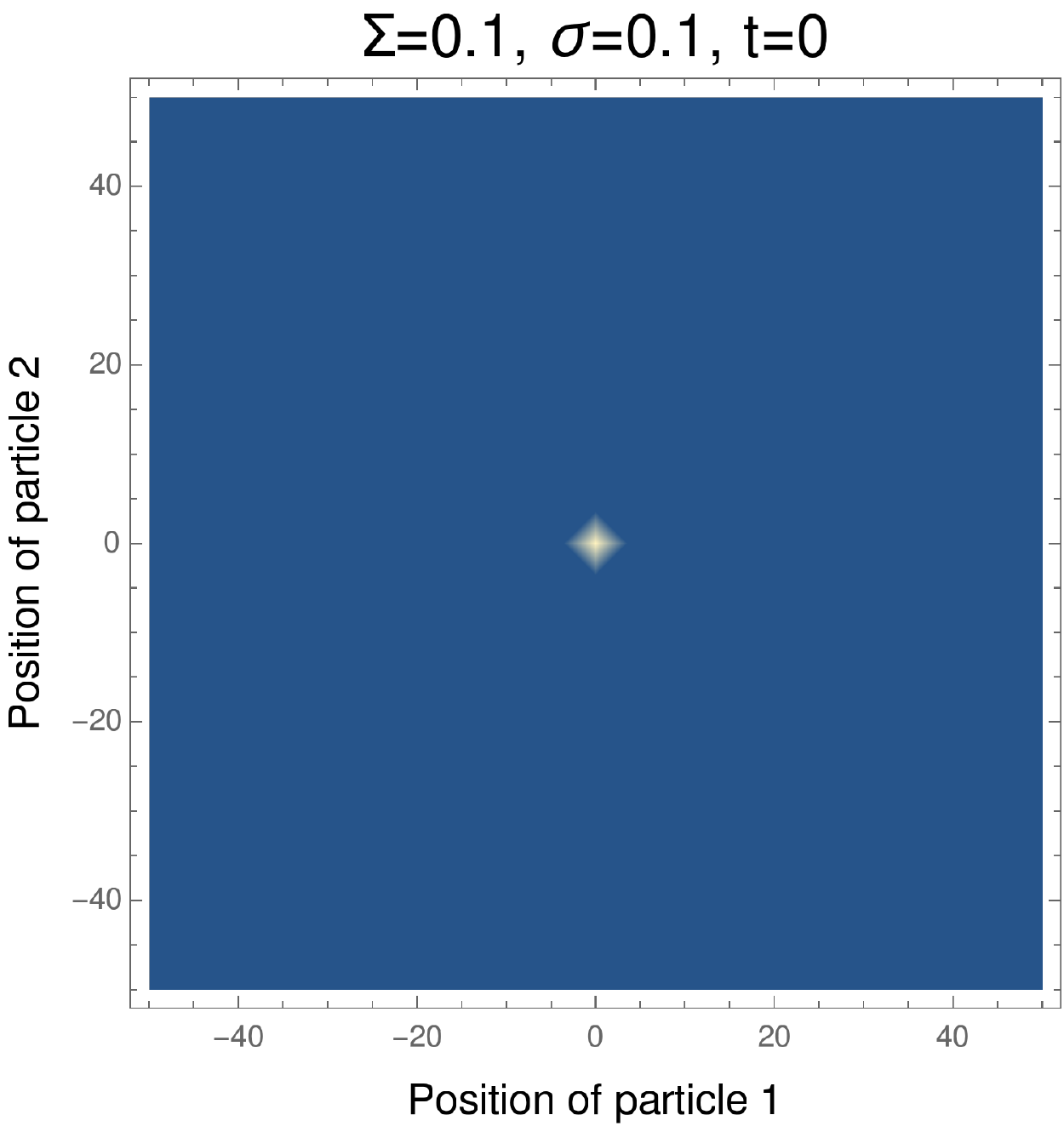}~\includegraphics[scale=0.32]{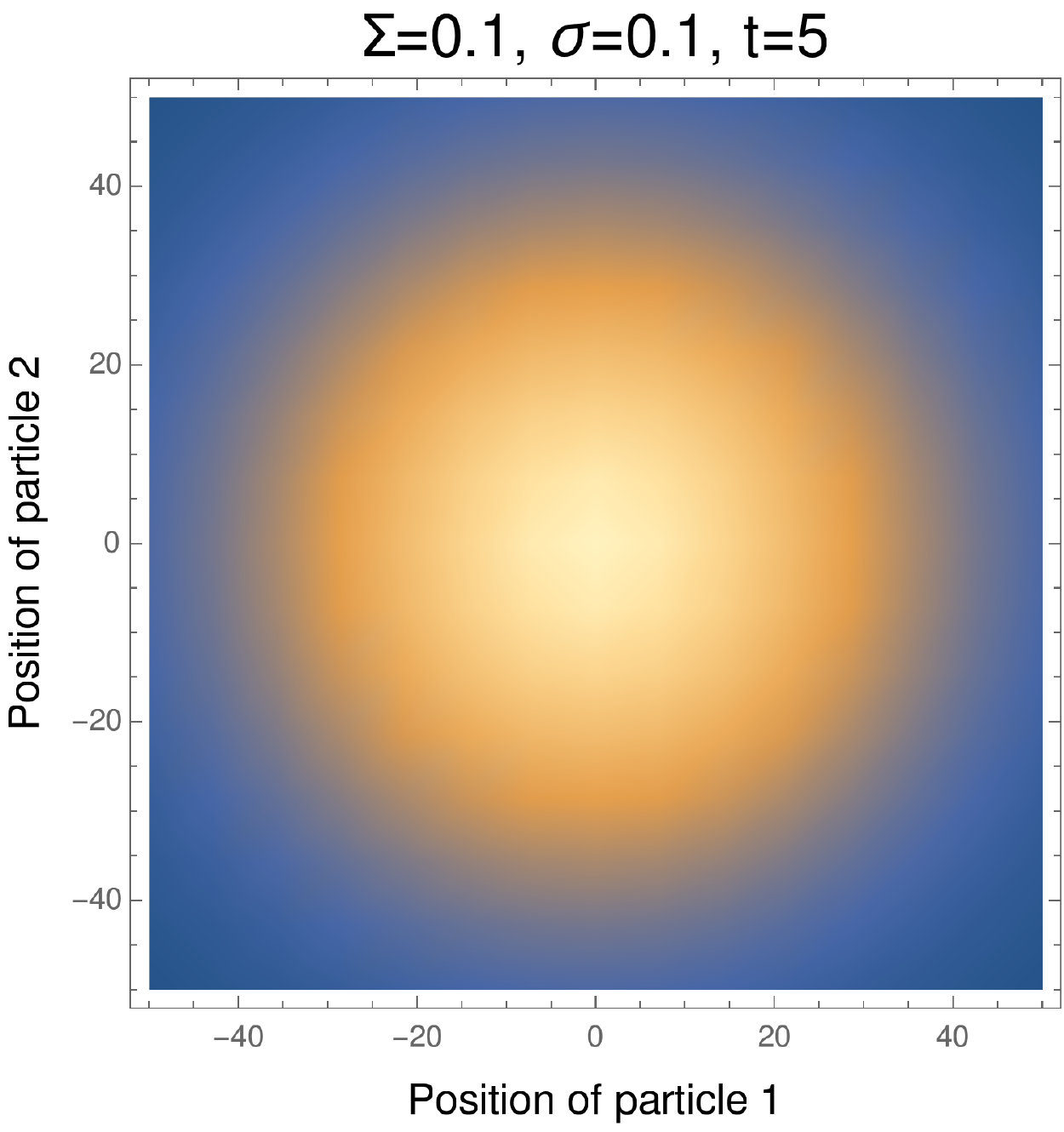}
\includegraphics[scale=0.32]{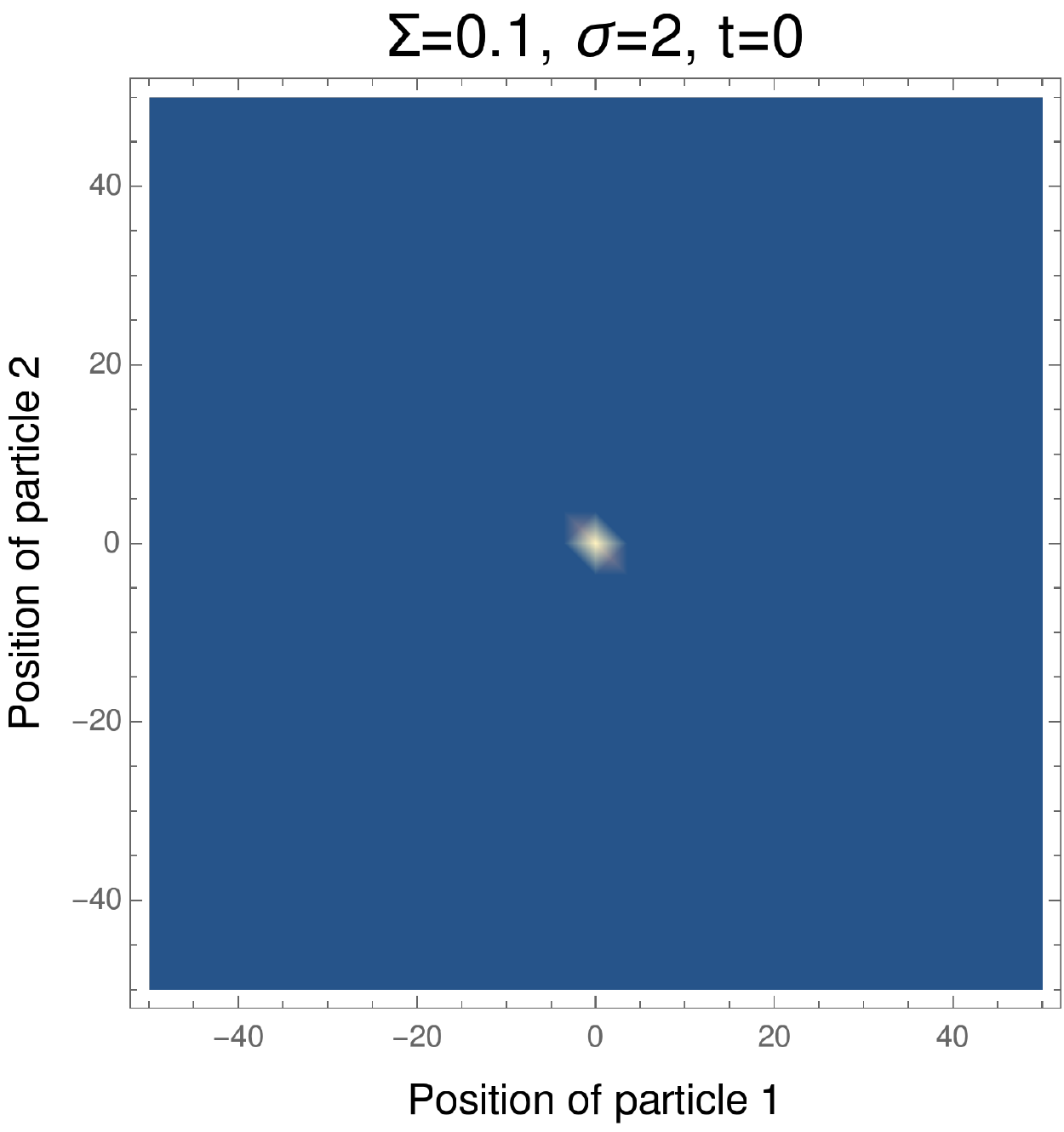}~\includegraphics[scale=0.32]{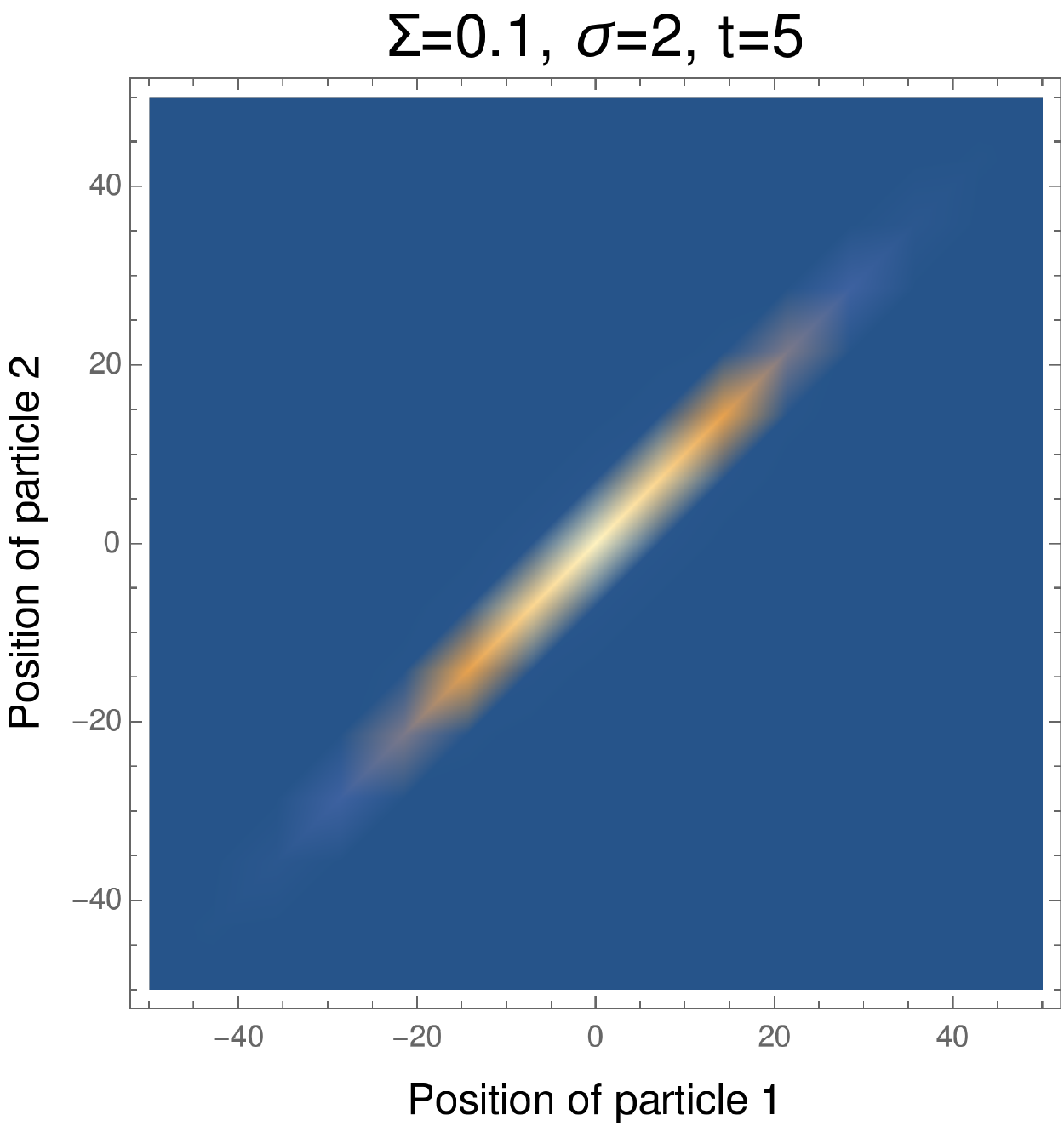}
\caption{Example evolution of a double Gaussian wave packet for a separable state (top) and an entangled one (bottom). We assume natural units ($\hbar = 1$ and $m=1$). \label{fig2}}
\end{figure}

Note, that if $\sigma = \Sigma$ the system is in a separable state ${\cal N}e^{-x_1^2/2\sigma^2}e^{-x_2^2/2\sigma^2}$. In this case the particles evolve independently and the system does not fulfil our composite particle criteria. However, the situation is different in the case of strong entanglement corresponding to $\Sigma \ll \sigma$. Since for long times the standard deviations of the centre of mass and of the relative position scale as $t/\Sigma$ and $t/\sigma$, respectively, we see that while the centre of mass gets delocalized, the distance between the particles does not change much (see Fig. \ref{fig2}).


It is useful to define $\tau \equiv \frac{m}{\hbar}\sigma^2$ as the lifetime of the composite particle and $\delta=\sigma/\sqrt{2}$ as its initial size. Intuitively, for $t>\tau$ the size of the composite particle is larger than $\sqrt{2}\delta$, which can be interpreted as a particle decay. Therefore, before the composite particle decays the centre of mass will spread over the distance $\Delta_{cm}(\tau)$, which can be evaluated with the help of (\ref{sigma}) and (\ref{purity})
\begin{equation}\label{entspread}
\Delta_{cm}(\tau) = \frac{1}{\sqrt{2}}\sqrt{\frac{\Sigma^4 + \sigma^4}{ \Sigma^2}}= \frac{\delta}{P}\sqrt{4-2P^2}.
\end{equation}
Since the purity $P$ measures the entanglement of pure states, we conclude that the value $\Delta_{cm}(\tau)$ is solely determined by the initial size of the composite particle and the entanglement of its constituents.
   
 
\subsection{Thermalization}

Next, we consider the model of thermalization of a Gaussian wave packet discussed in \cite{GaussTerm}. This time, each particle has a momentum $k_1$ and $k_2$ so the initial wave-function is given by 
\begin{eqnarray}
\psi_{k_1,k_2}(x_1,x_2,t=0) &=& {\cal N}e^{-\frac{(x_1-x_2)^2}{4\sigma^2}+i\frac{(k_1 - k_2)(x_1 - x_2)}{2}} \nonumber \\
&\times & e^{-\frac{(x_1+x_2)^2}{4\Sigma^2}+i\frac{(k_1 + k_2)(x_1 + x_2)}{2}}.
\end{eqnarray} 
However, in a thermal state momenta of particles are random and are described by some probability distribution
\begin{equation}
\rho(x_1,x_2,t) = \int dk_1 dk_2 \mu(k_1)\nu(k_2)|\psi_{k_1,k_2}(x_1,x_2,t)|^2,
\end{equation} 
where $\mu(k_1)$ and $\nu(k_2)$ are the distribution of respective momenta.

\begin{figure}
\includegraphics[scale=0.32]{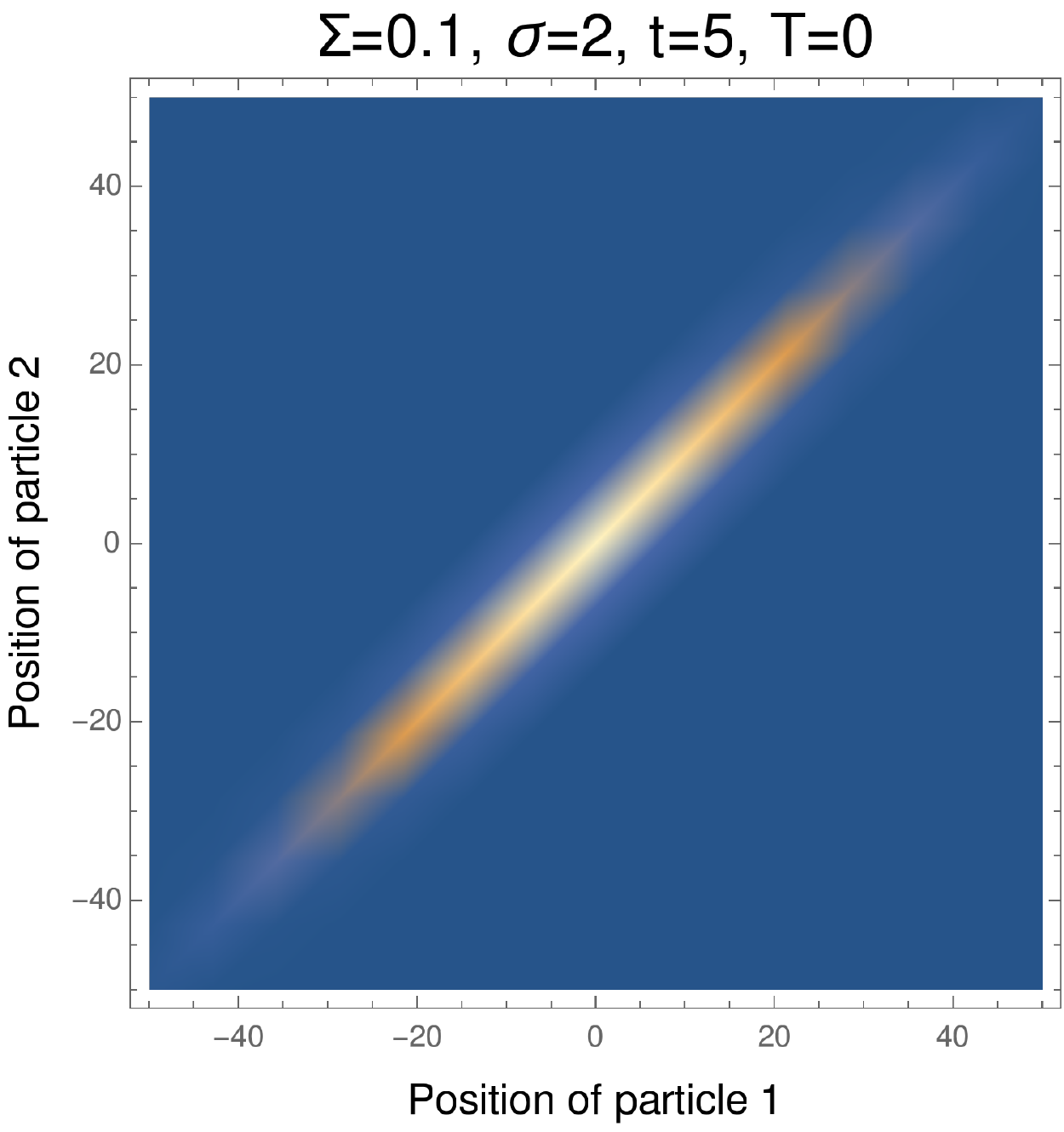}~\includegraphics[scale=0.32]{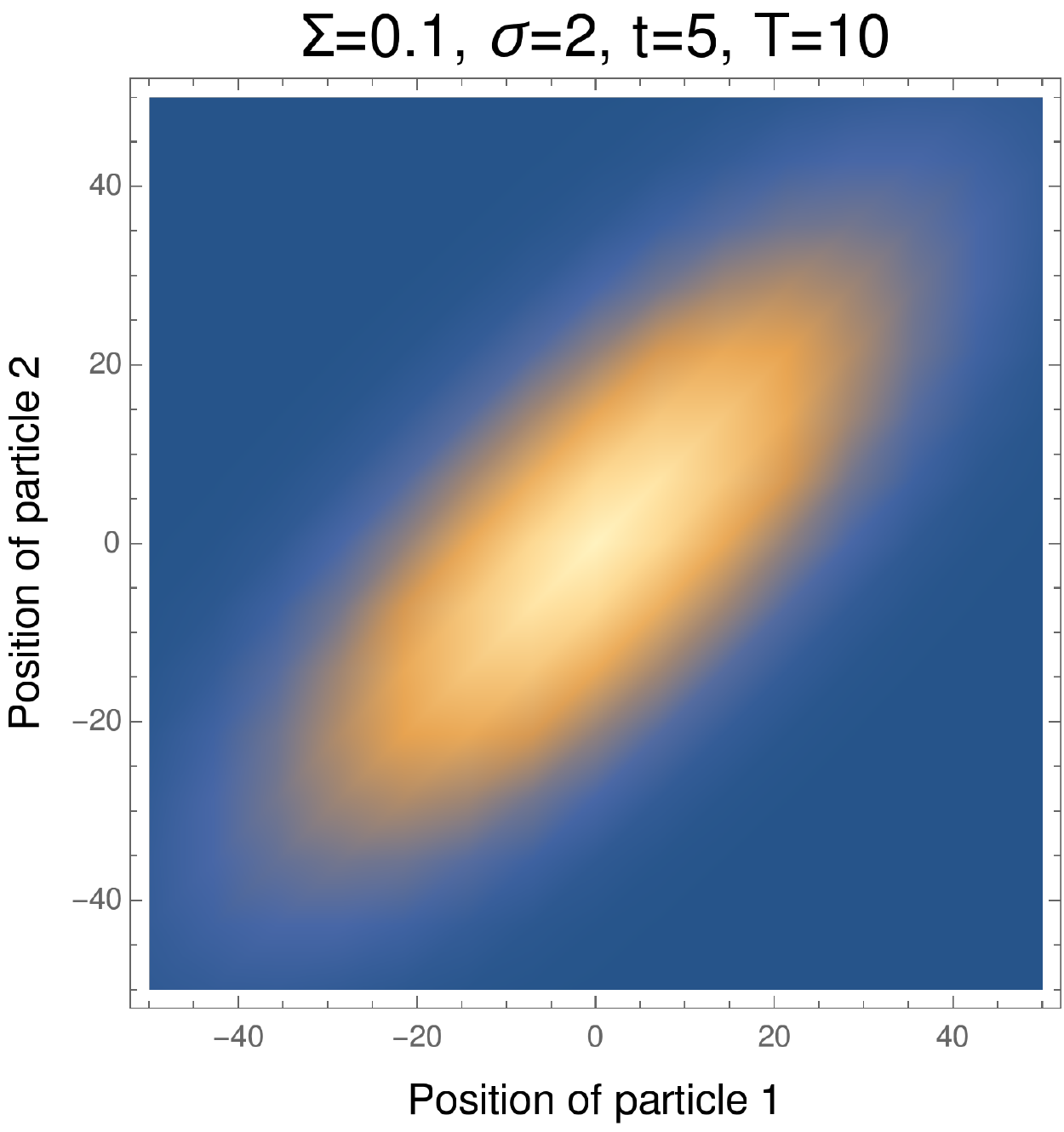}
\includegraphics[scale=0.32]{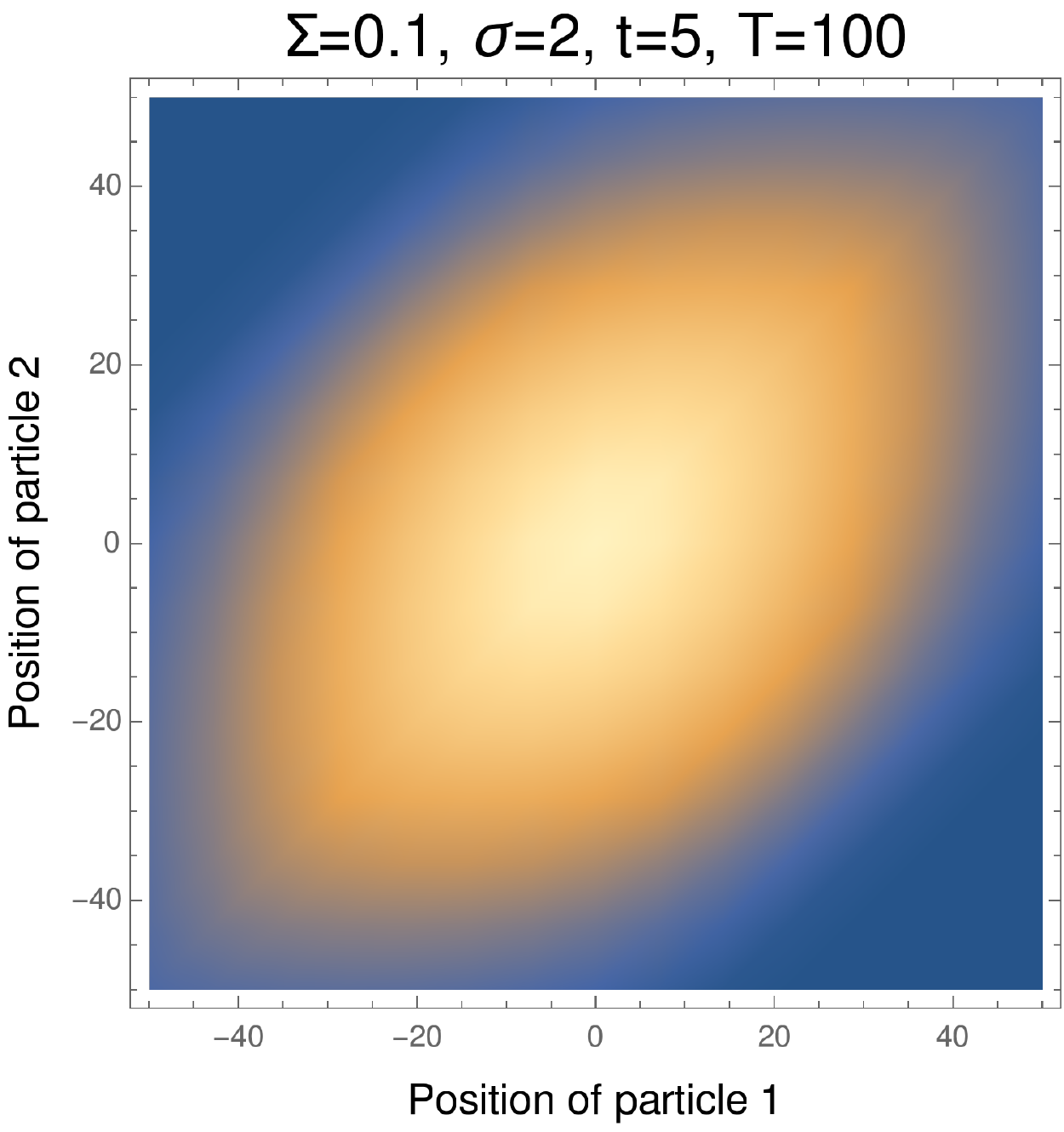}~\includegraphics[scale=0.32]{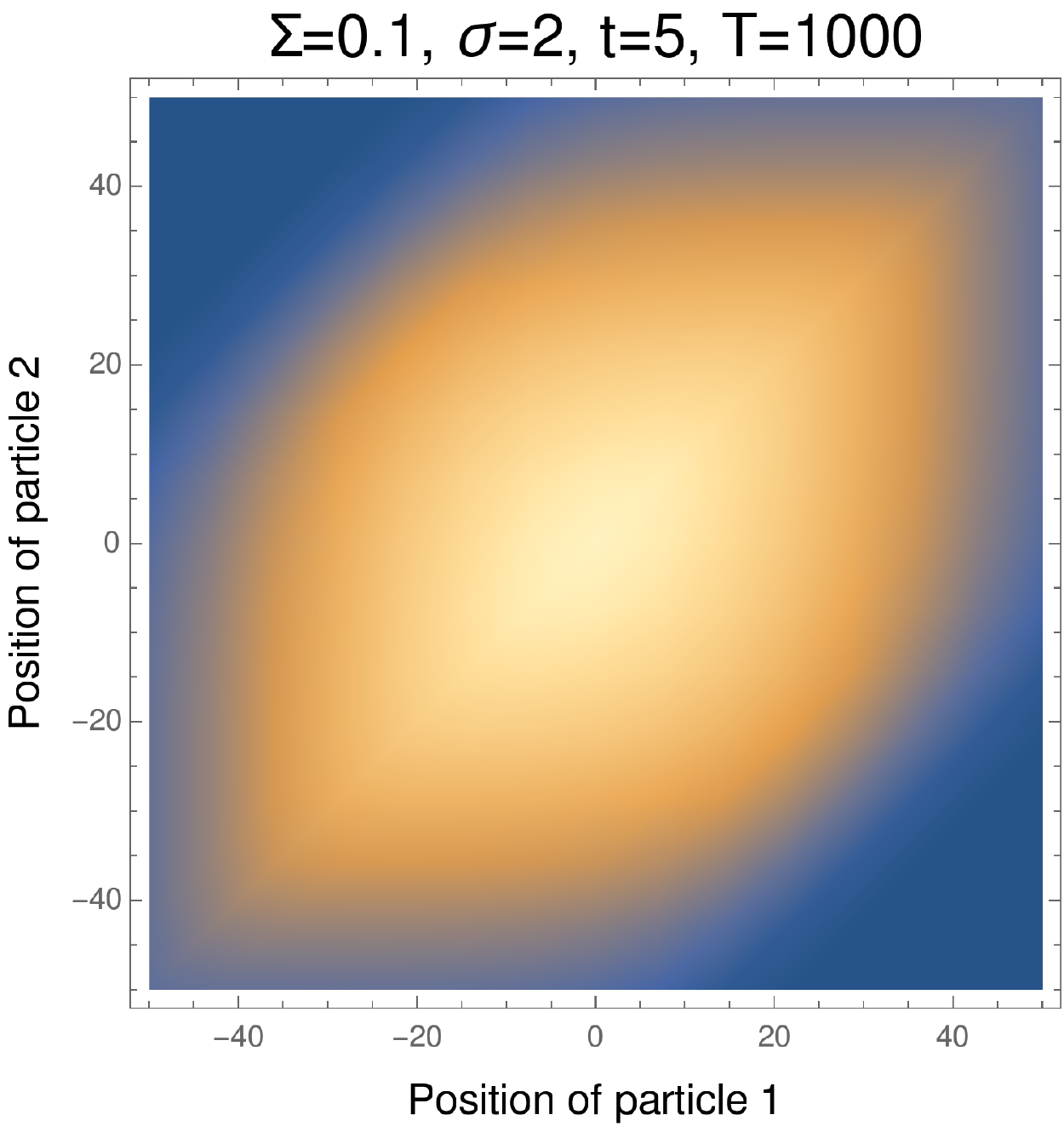}
\caption{Thermalization of a double Gaussian wave packet for various temperatures. We used natural units $k_B=1$ and $\hbar=1$. \label{fig3}}
\end{figure}

We assume that $k_1$ and $k_2$ are discrete and that they are independent and identically distributed according to Maxwell distribution
\begin{equation}
\mu(k,T)=\frac{1}{Z} e^{-\frac{\hbar^2 k^2}{2 k_B T}},
\end{equation}
where $Z=\sum_k e^{-\frac{\hbar^2 k^2}{2 k_B T}}$, $T$ is the temperature, and $k_B$ is the Boltzmann constant. In Fig. \ref{fig3} we present the effects of thermalization for different temperatures. We assumed that each momentum can take value $\pm\frac{\hbar n\pi}{5}$, where $n=0,1,\ldots,10$. As expected, thermalization leads to a decay of a composite particle. 


Note, that the thermalization affects rather the relative position, than the position of the centre of mass. Let $\Delta_T(k)$ be the thermal spread of the momentum. Therefore, due to thermal fluctuations, after time $t$ the particle of mass $m$ is spread over the distance $\frac{\Delta_T(k)}{m} t$. In our case $\Delta_T(k) \leq 4\hbar\pi$. On the other hand, as argued above, the natural spreading of the centre of mass is approximated as $\approx \frac{\hbar}{\sqrt{2}m\Sigma} t$. Therefore, the thermal effects would dominate the spreading of the centre of mass if $\frac{\hbar}{\sqrt{2}\Sigma} \ll \Delta_T(k)$. 

Since we consider wave packets for which $\Sigma \ll \sigma$, it is justified to approximate that in our case the thermalization affects only the relative position. In this case, we can use the result from \cite{GaussTerm} which implies that thermalization affects the standard deviation of the relative position in the following way
\begin{equation}\label{sigmaT}
\Delta_T (x_1 - x_2) (t) = \frac{1}{\sqrt{2}}\sqrt{\sigma^2 + \left(\frac{\hbar^2}{m^2 \sigma^2} + \frac{k_B T}{m}\right)t^2}.
\end{equation}
Therefore, for large temperatures the time of the composite particle decay can be approximated as $\tau_{T} \approx \sqrt{\frac{m}{k_B T}}$, in which case the initial entanglement does not matter anymore. The temperatures for which the thermal component is smaller than the original one correspond to $\frac{\hbar^{2}}{m k_B \sigma^2} > T$. For example, for a Cooper-like pair of two electrons that are one hundred nanometers from each other ($10^{-7}$ m) we get $T < 0.088 K$.   

		
\section{Oscillations of entangled particles}

We observed that in certain conditions the free evolution of an entangled pair can be interpreted as the free evolution of a single quantum particle. It is therefore natural to speculate that entangled pairs will manifest other types of single-particle quantum behaviour. In this section we show that this speculation is only partially true. While entangled particles stay together throughout the evolution, natural measurements on such systems can only detect elementary oscillations of the constituents. More precisely, measurements which are not post-selective and which do not reveal the internal structure of the composite particle are not capable of detecting composite oscillations. 

A bipartite composite particle should exhibit a wavelike behaviour with the corresponding wavelength equal to the half of the single-particle wavelength. This is because the de Broglie wavelength of an object is inversely proportional to momentum $\lambda_0=h/p$. Two particles, each having momentum $p$, have a joint momentum of $2p$, therefore the collective de Broglie wavelength should be equal to $\lambda_0/2$. This phenomenon has practical applications and can be exploited in quantum metrology \cite{QuantInterf}. However, we show that the fractional wavelength could be only observed in situations in which particles interact or measurements are post-selective and address the internal structure of the composite particle. This will be explained in more details in a moment. 


\subsection{Discrete double Gaussian state}

Once again we consider evolutions of the double Gaussian state, however this time, for the purpose of numerical simulations, we assume that the space is finite and discrete. Therefore, the state is given by  
\begin{equation}\label{gauss2}
|\psi \rangle =  \mathcal{N}  \sum_{\substack{x_1 , x_2 = 1 }}^{d} e^{\frac{(x_1 + x_2)^2}{4\sigma ^2}}e^{\frac{(x_1 - x_2)^2}{4\Sigma ^2}} a^{\dagger}_{x_1} b^{\dagger}_{x_2}|0\rangle,
\end{equation} 
where  $a^{\dagger}_{x_1}$ creates the first particle in position $x_1$ and $b^{\dagger}_{x_2}$ creates the second one in position $x_2$. Moreover, we assume periodic boundary conditions, i.e., $x_j+d \equiv x_j$ for $j=1,2$. The evolution will be generated by $H = H_{free} + V$, where
\begin{equation}\label{Hfree}
H_{free} = - \sum_{x=1}^d \left( a^{\dagger}_{x+1} a_x +  a^{\dagger}_{x} a_{x+1}+b^{\dagger}_{x+1} b_x +  b^{\dagger}_{x} b_{x+1}\right)
\end{equation} 
is the free evolution Hamiltonian that generates hopping between neighbouring lattice points. The second term $V$ corresponds to a potential that will change from case to case.


\subsection{Measurements}

Before we proceed to study the dynamics, let us discuss in more details the role of measurements in our scenario. As we already stated, the constituents need to be close to each other. Therefore, the composite particle cannot produce two clicks at two spatially separated detectors. However, this is not a precise statement, since we need to clarify the meaning of {\it close} and {\it spatially separated}.

Note, that the definition of the double Gaussian state implies that the two particles are not in the same place. The standard deviation of their relative position is given by $\delta=\sigma/\sqrt{2}$, which we chose to interpret as the size of the composite particle. Any single-partite treatment of a fundamentally composite system is based on some kind of ignorance. In our case, it is the ignorance of the internal structure. More precisely, we focus on a centre of mass and ignore the relative position, as long as the distance between the constituents is not larger than some critical value $\Delta$ ($\Delta \geq \delta$). This critical value leads to an effective coarse graining of space. 

In addition, the two particles in a state represented by Eq. (\ref{gauss2}) are described by different creation operators. However, from the observer point of view they should not be distinguishable, since the ability to distinguish them would automatically imply that the investigated object has an internal structure and that the constituents can be individually addressed. Therefore, the measured observables should be symmetric under permutation of particles and should take into account the coarse graining of space.

As we noted before, we consider a discrete space with $d$ positions and with periodic boundary conditions. We set $d=m\Delta$, where $\Delta$ is the size of the coarse grained unit cell and $m$ is the number of such cells. We choose the following operators to describe detectors in our scenarios
\begin{equation}\label{detector}
D_j = \sum_{x_1,x_2=j\Delta + 1}^{(j+1)\Delta} a_{x_1}^{\dagger}a_{x_1} b_{x_2}^{\dagger}b_{x_2},
\end{equation}
where $j=0,\ldots,m-1$ labels the coarse grained unit cells. For a single particle of type $a$ and a single particle of type $b$ the above operator has an eigenvalue $1$, if both particles are in the same coarse grained unit cell $j\Delta + 1 \leq x_1,x_2 \leq (j+1)\Delta$, and an eigenvalue $0$ otherwise. In simple words, there are $m$ detectors, each corresponding to a different coarse grained cell, and at most one of them can click -- register an outcome $1$. If the composite particle falls apart, i.e., one particle is in one cell $j\Delta + 1 \leq x_1 \leq (j+1)\Delta$ and the other in some other cell $j'\Delta + 1 \leq x_2 \leq (j'+1)\Delta$ ($j\neq j'$), then none of the detectors click. Note, that coarse graining leads to some imperfections, namely it can happen that the two particles are closer than $\Delta$, but they are still in two different cells. For example, $x_1 = \Delta - 1$ whereas $x_2 = \Delta + 1$ in which case the first particle is in the cell corresponding to $D_0$ and the second one in the cell corresponding to $D_1$. Still, the idea is to take $\Delta$ large enough to assure that such possibilities occur with low probabilities. 

Finally, note that we could choose a different detection operator, like 
\begin{equation}
D^{(a)}_j + D^{(b)}_j = \sum_{x_1=j\Delta + 1}^{(j+1)\Delta} a_{x_1}^{\dagger}a_{x_1}  + \sum_{x_2=j\Delta + 1}^{(j+1)\Delta} b_{x_2}^{\dagger}b_{x_2}.
\end{equation}
Such operator would detect each particle separately (without distinguishing which is which) and would be able to say where the particles are after the composite particle decayed. However, our goal is to study the effects of entanglement, whereas the average value of the above operator does not give any information about the correlations between the particles since
\begin{equation}
\langle D^{(a)}_j + D^{(b)}_j \rangle = \text{Tr}\{D^{(a)}_j \rho_a \} + \text{Tr}\{D^{(b)}_j \rho_b \},
\end{equation}
where $\rho_a$ and $\rho_b$ are the reduced density matrices of particles $a$ and $b$, respectively. Only the variance, or the higher moments, of the above operator can reveal correlations between the particles.


\subsection{Mach-Zehnder-like setup}

\begin{figure}
\includegraphics[scale=0.45]{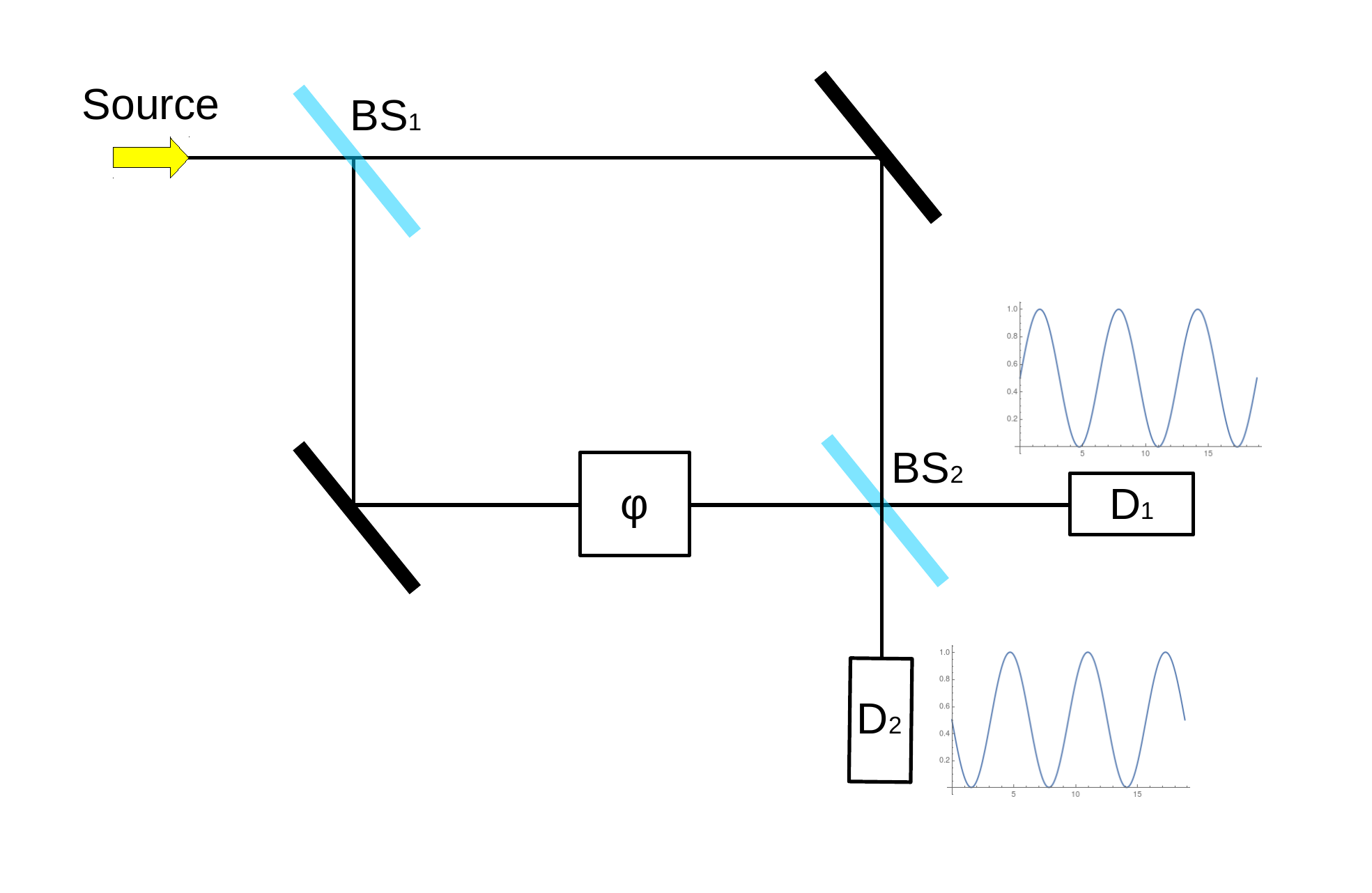}
\vspace{-10mm}
\caption{Schematic representation of Mach-Zehnder interferometer. \label{fig4}}
\end{figure}

It was proposed in \cite{PhotonicdeBroglieWaves95} that wave-like properties of composite particles can be observed in Mach-Zehnder Interferometer (MZI) by detecting an interference pattern corresponding to a collective de Broglie wavelength. This avenue of research was further explored by other research groups \cite{Fonseca99,Sackett00,Fonseca01,Edamatsu02}. It is also important to mention that collective de Broglie wavelengths are a central subject of quantum metrology \cite{Dowling} which exploits them to exceed the classical limits in imaging and sensing. All of these investigations are based on either interacting particles, or specially prepared nonclassical states (like the NOON states \cite{Dowling}) and measurement setups that take into account only some special intra-particle properties (internal structure of composite particles in our language).

In a standard single-particle MZI experiment the particle goes through a setup presented in Fig. \ref{fig4}. After the first beam splitter (BS$_1$) the particle is in a superposition of travelling along one of two paths. The length of one path can be extended and as a result the particle travelling on this path acquires an extra phase factor $e^{i\varphi}$. 
The interference pattern at the outputs after the second beam splitter (BS$_2$) is detected by detectors $D_1$ and $D_2$. The amplitude at $D_1$ results from the superposition of amplitudes of two events $\{r,t\}$ and $\{t,r\}$, where $\{r,t\}$ means reflecting from BS$_1$ and transmitting through BS$_2$. The amplitude at $D_2$ results from the superposition of $\{r,r\}$ and $\{t,t\}$. If we assume $50/50$ BS and use the convention that the reflection causes a phase shift of $i$, then the resulting amplitude at detector $D_1$ is $\frac{i}{2}(1+e^{i\varphi})$, whereas the one at $D_2$ is $\frac{1}{2}(1-e^{i\varphi})$. We see that both amplitudes are periodic functions of $\varphi$ and the periods of oscillations are $2\pi$.    

Next, consider two particles entering the MZI together through the same port. First, assume that these particles are independent and non-interacting. There are four detection events (assuming no losses): $D_1 \times D_1$, $D_1 \times D_2$, $D_2 \times D_1$, and $D_2 \times D_2$. Here $D_1 \times D_2$ means that the first particle is detected at $D_1$ and the second at $D_2$. One can find events leading to the corresponding detection events. For example, the events leading to $D_1 \times D_1$ are $\{r,t\}\times \{r,t\}$, $\{r,t\}\times \{t,r\}$, $\{t,r\}\times \{r,t\}$ and $\{t,r\}\times \{t,r\}$. Assuming $50/50$ BS, the amplitude of $D_1 \times D_1$ is $-\frac{1}{4}(1+2e^{i\varphi}+e^{i2\varphi})$, of $D_1 \times D_2$ and $D_2 \times D_1$ is $\frac{i}{4}(1-e^{i2\varphi})$, and of $D_2 \times D_2$ is $\frac{1}{4}(1-2e^{i\varphi}+e^{i2\varphi})$. We see, that the amplitudes of $D_1 \times D_2$ and $D_2 \times D_1$ have double oscillation periods equal to $\pi$. However, in order to detect them one needs to register clicks at two different places, which does not meet our definition of a particle. Moreover, to stress the lack of a composite particle-like nature of events $D_1 \times D_2$ and $D_2 \times D_1$, note that the above scenario can be considered in a spatially separated setup in which each particle enters a different MZI (similar to the one studied in \cite{Fonseca01}). 

Finally, consider once more two particles in the MZI, however this time we assume that the particles always go together (either due to interaction or due to any other mechanism). Therefore, events in which particles go on different paths, like $\{r,t\}\times \{t,r\}$, are not possible. As a result, the amplitude of event $D_1 \times D_1$ is a superposition of amplitudes of events $\{r,t\}\times \{r,t\}$ and $\{t,r\}\times \{t,r\}$, which gives $-\frac{1}{4}(1+e^{i2\varphi})$. This amplitude has a period of $\pi$, therefore in this case it is possible to detect a collective de Broglie wavelength in an event that meets our composite particle definition, i.e., in an event in which both particles are in the same place.  

We see that the crucial effect responsible for the observation of the collective de Broglie wavelength is that all particles making up a composite system stay together, i.e., they collectively reflect or go through BS. In a simple two-mode BS this is not possible without the intra-particle interaction, or some kind of post selection (ignoring events in which particles separated). However, in the previous section we showed that entangled non-interacting particles can stay together while their centre of mass gets delocalized. Therefore, it is natural to expect that in generalized MZI experiments such systems will produce interference patterns corresponding to collective de Broglie wavelengths, obeying our composite particle definition and without a need to examine the relative position of particles (the internal structure of the composite particle). We investigate this hypothesis below.  

The MZI experiment consists of five stages: (I) state preparation, (II) splitting of a wave-packet into two regions, (III) phase shift in the second region, (IV) recombination of a wave-packet in a single region, (V) measurement. We implement the MZI-like setup in our system by dividing the space into four regions, i.e., $\Delta=d/4$. As a result, we obtain four coarse grained cells $j=0,1,2,3$. It may seem strange that we decided to use four cells, instead of two (like in a standard MZI). However, if we used two cells and prepared the wave packet in the centre of one cell, then after the splitting the two wave packets would lie on the boundaries of both cells (which leads to the problems mentioned above). Therefore, we used four cells, because after the splitting the left and the right-going wave packets are at the centres of the neighbouring cells.

(I) We prepare the double Gaussian state centred in one of the cells, say $j=1$ (the centre of this cell corresponds to $3d/8$)
\begin{equation}
|\psi \rangle =  \mathcal{N}  \sum_{\substack{x_1 , x_2 = 1 }}^{d} e^{\frac{(x_1 + x_2 - 3d/4)^2}{4\sigma ^2}}e^{\frac{(x_1 - x_2)^2}{4\Sigma ^2}} a^{\dagger}_{x_1} b^{\dagger}_{x_2}|0\rangle.
\end{equation}
Note, that the perfect Gaussian state is extended over whole space from $x=-\infty$ to $x=+\infty$. Here, due to the finiteness of space, the wave packet is only supported on $d$ positions. This is not a big problem, since we demand that almost all packet is initially localized in one cell. 

\begin{figure}
\includegraphics[scale=0.22]{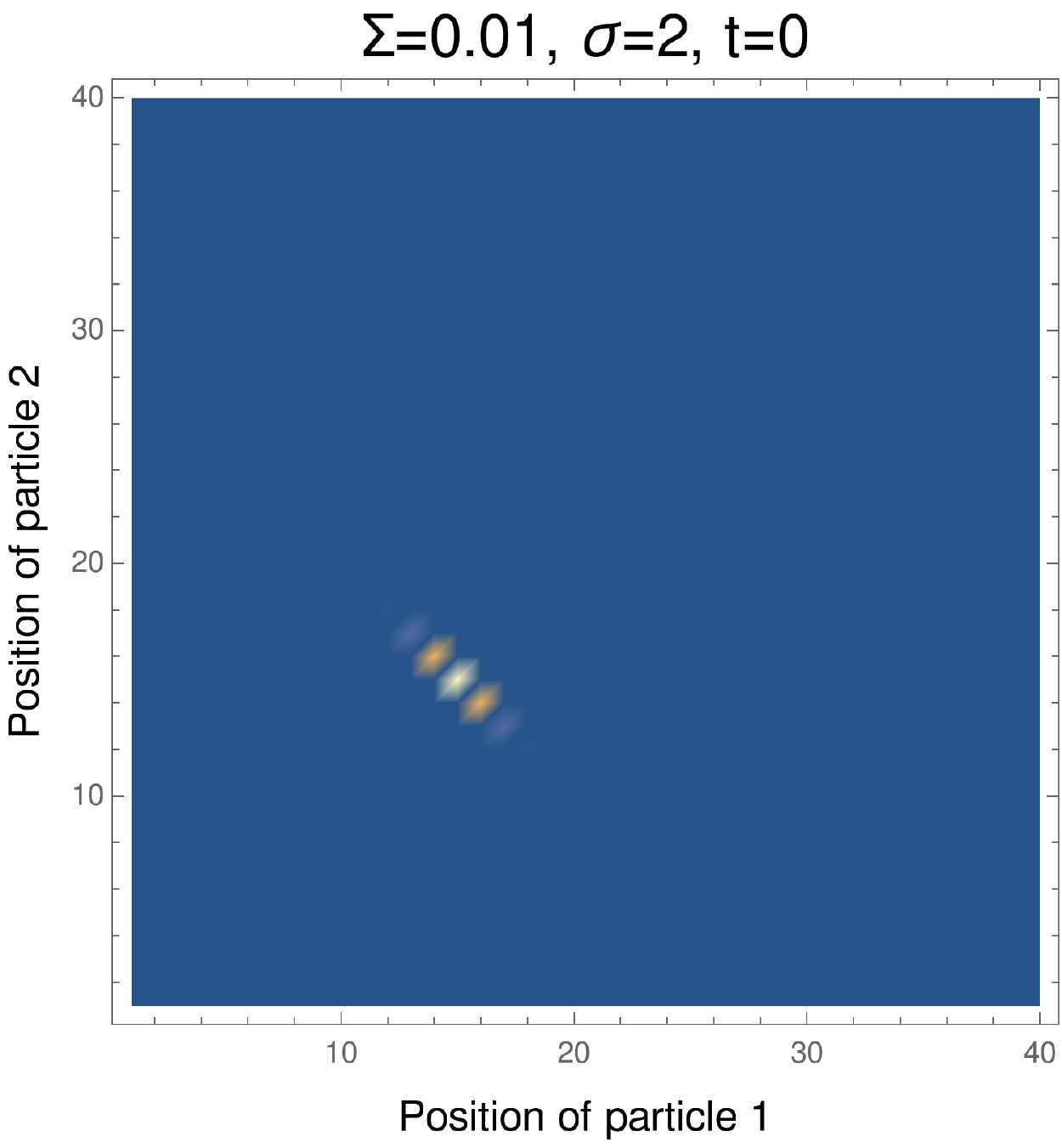}~\includegraphics[scale=0.22]{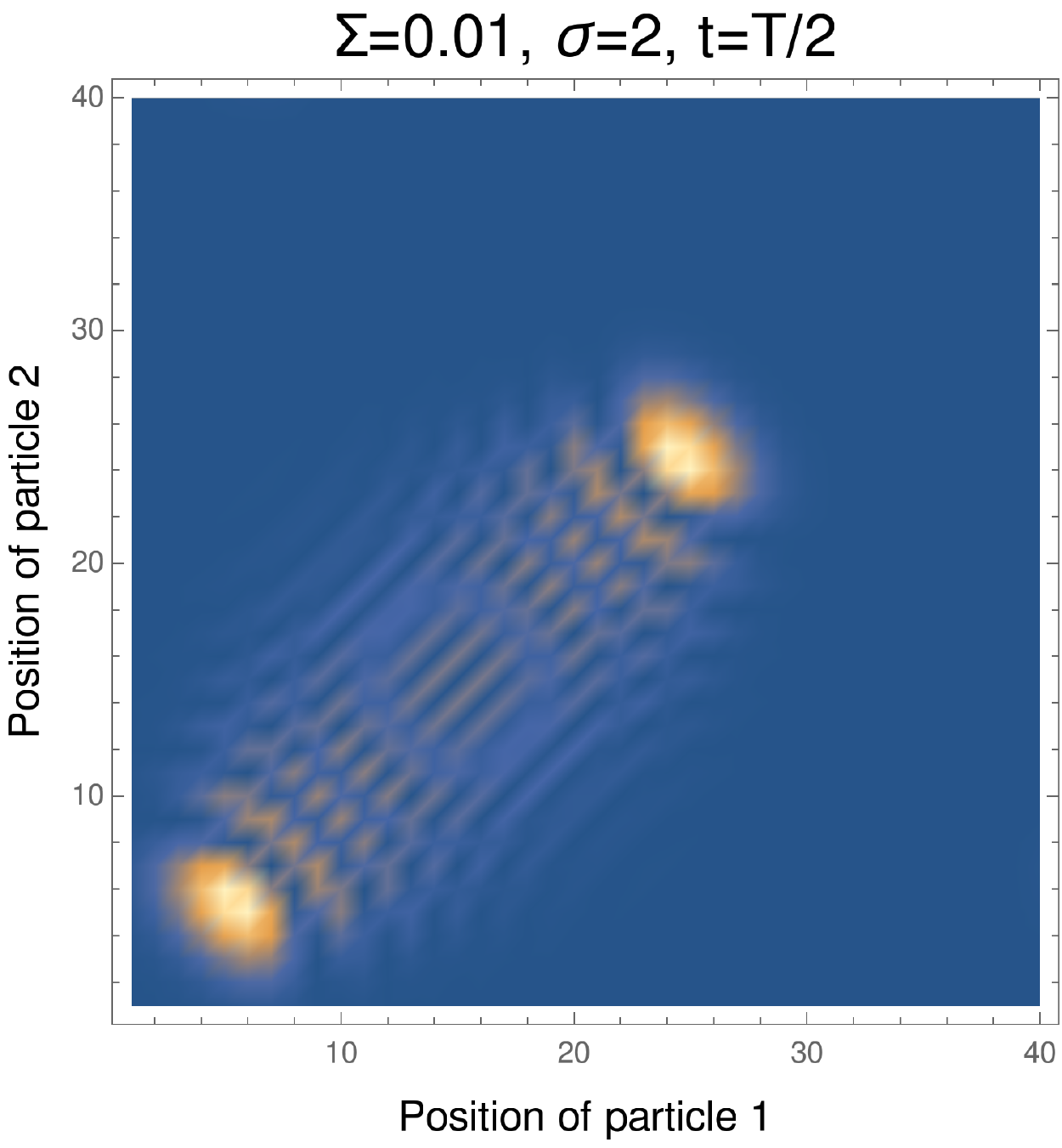}~\includegraphics[scale=0.22]{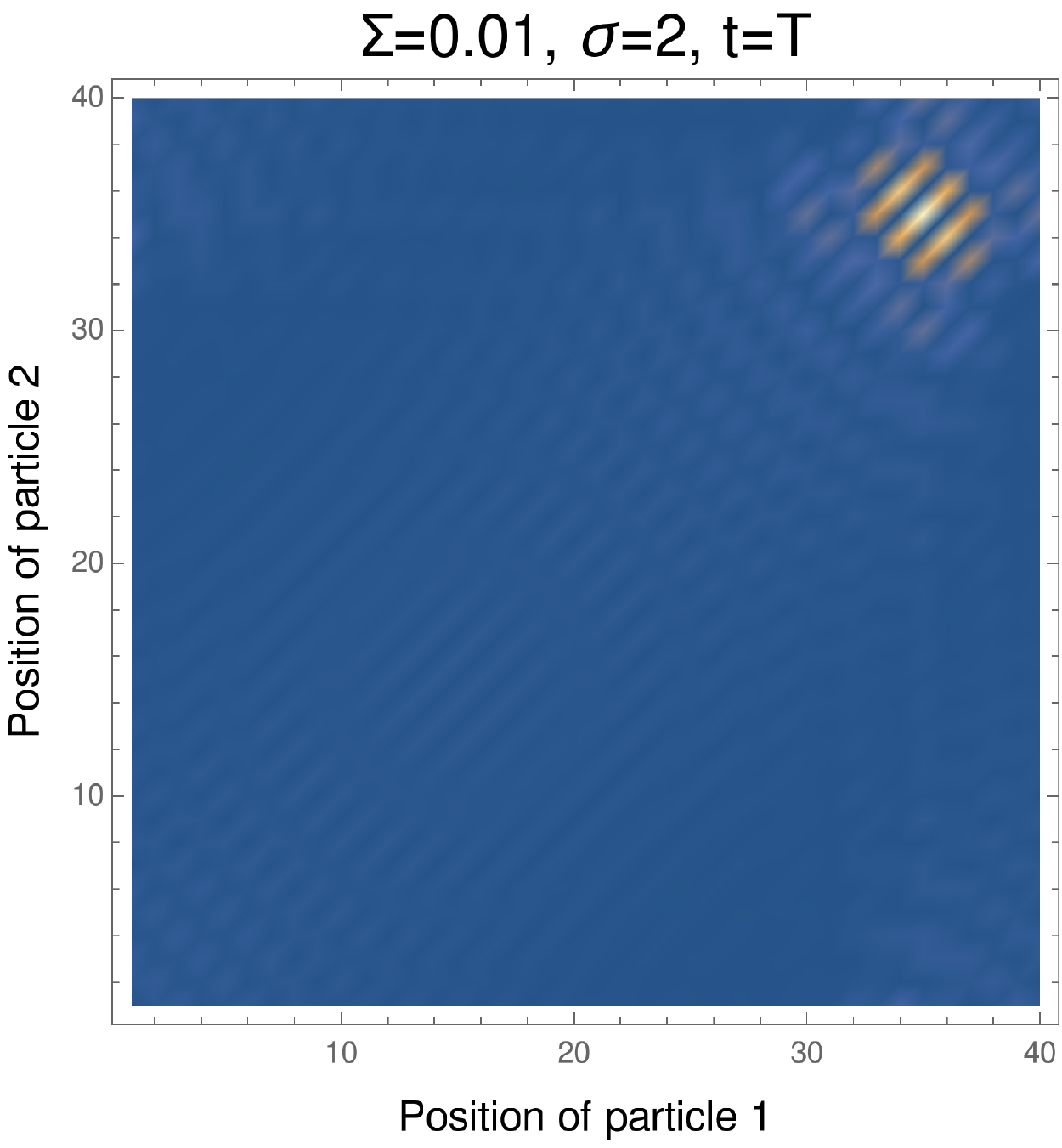}
\includegraphics[scale=0.22]{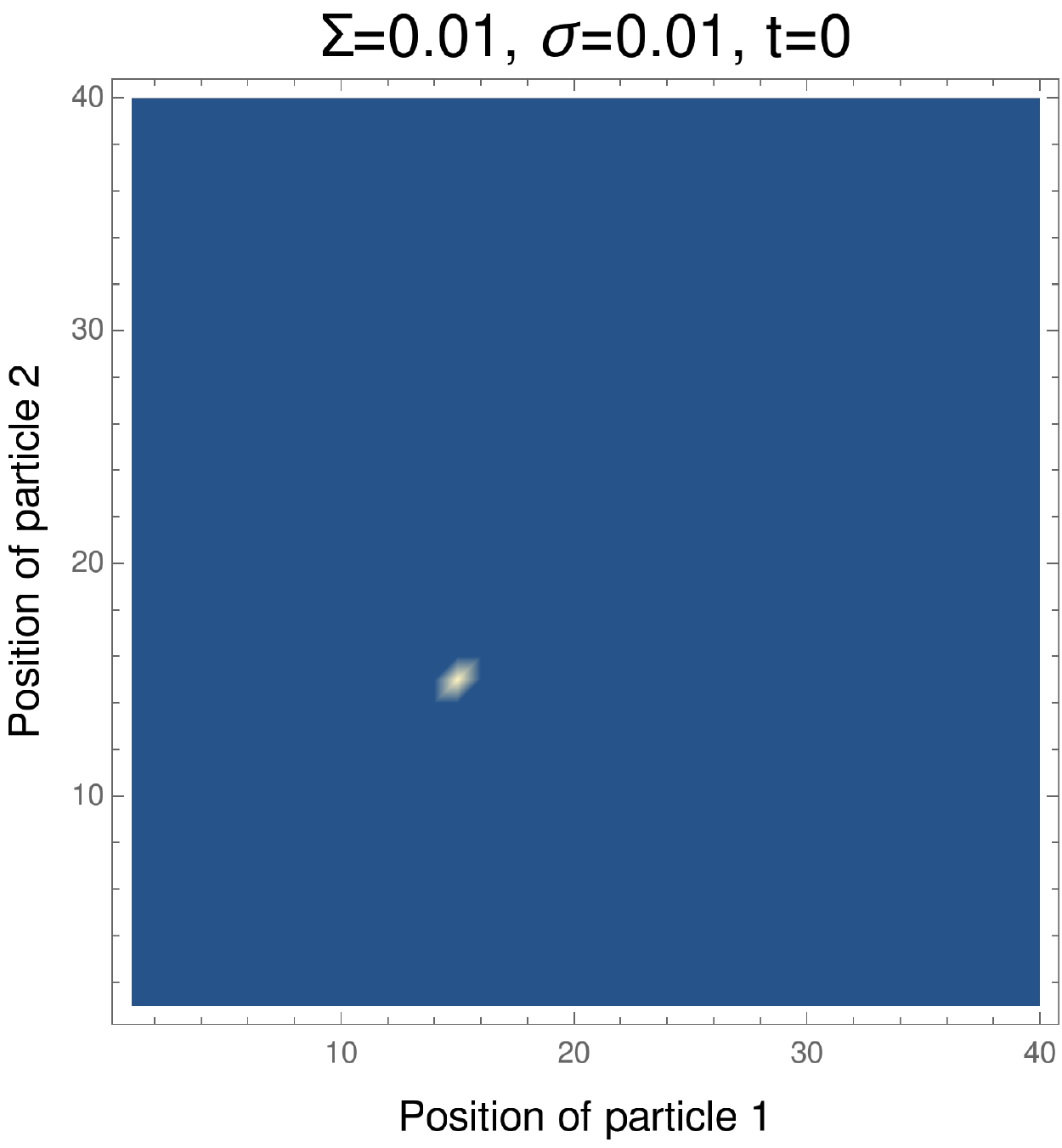}~\includegraphics[scale=0.22]{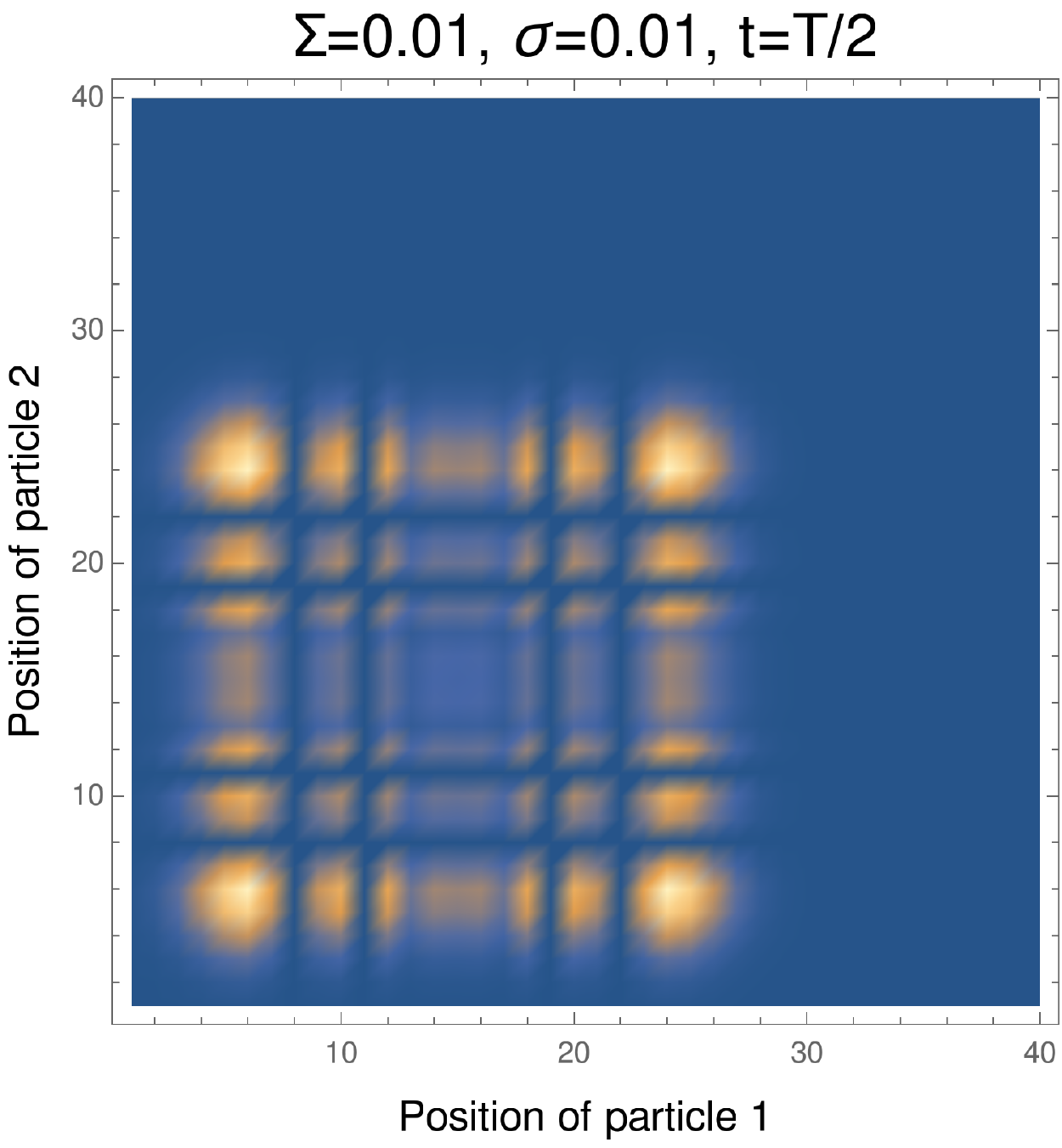}~\includegraphics[scale=0.22]{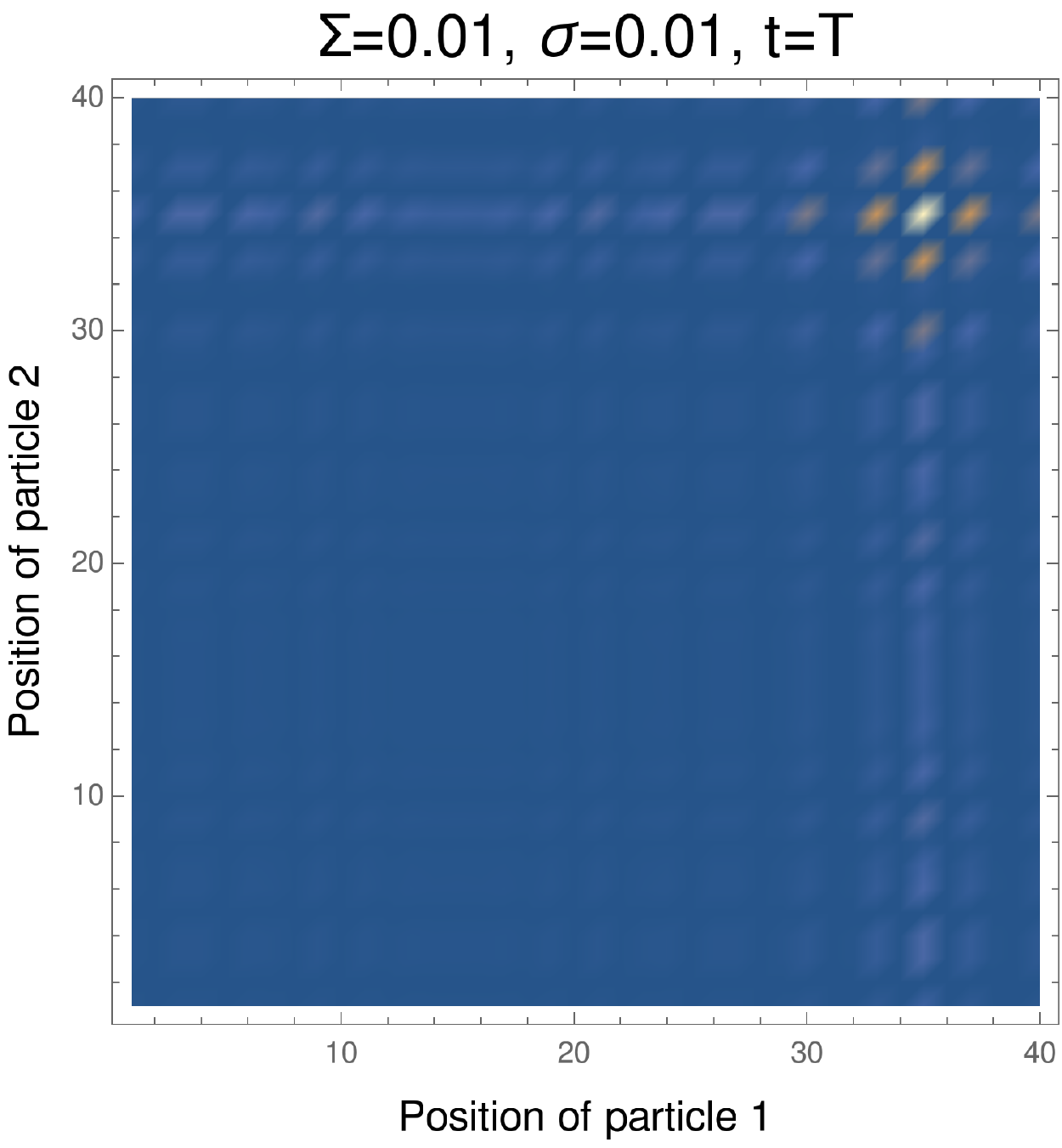}
\includegraphics[scale=0.22]{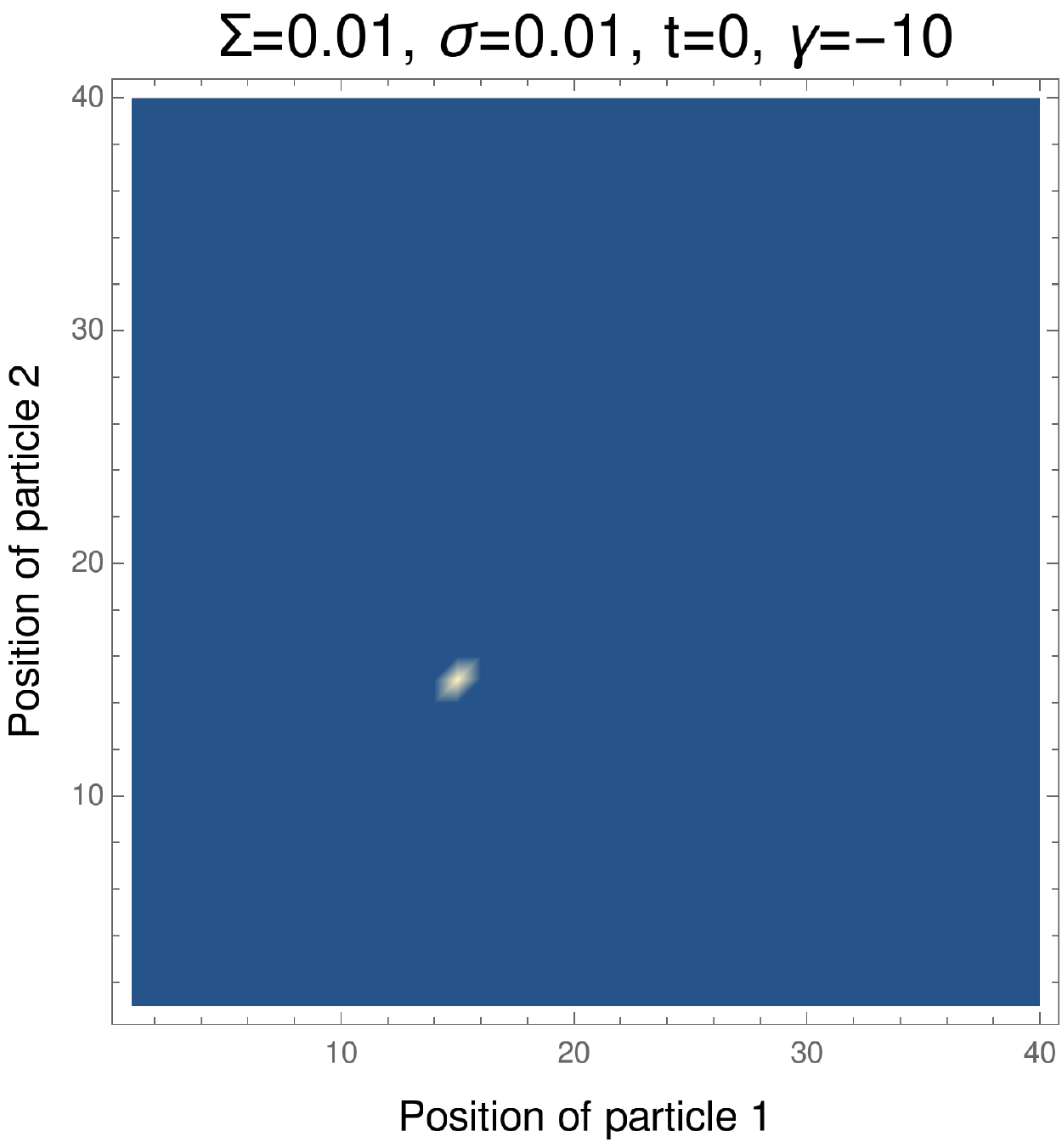}~\includegraphics[scale=0.22]{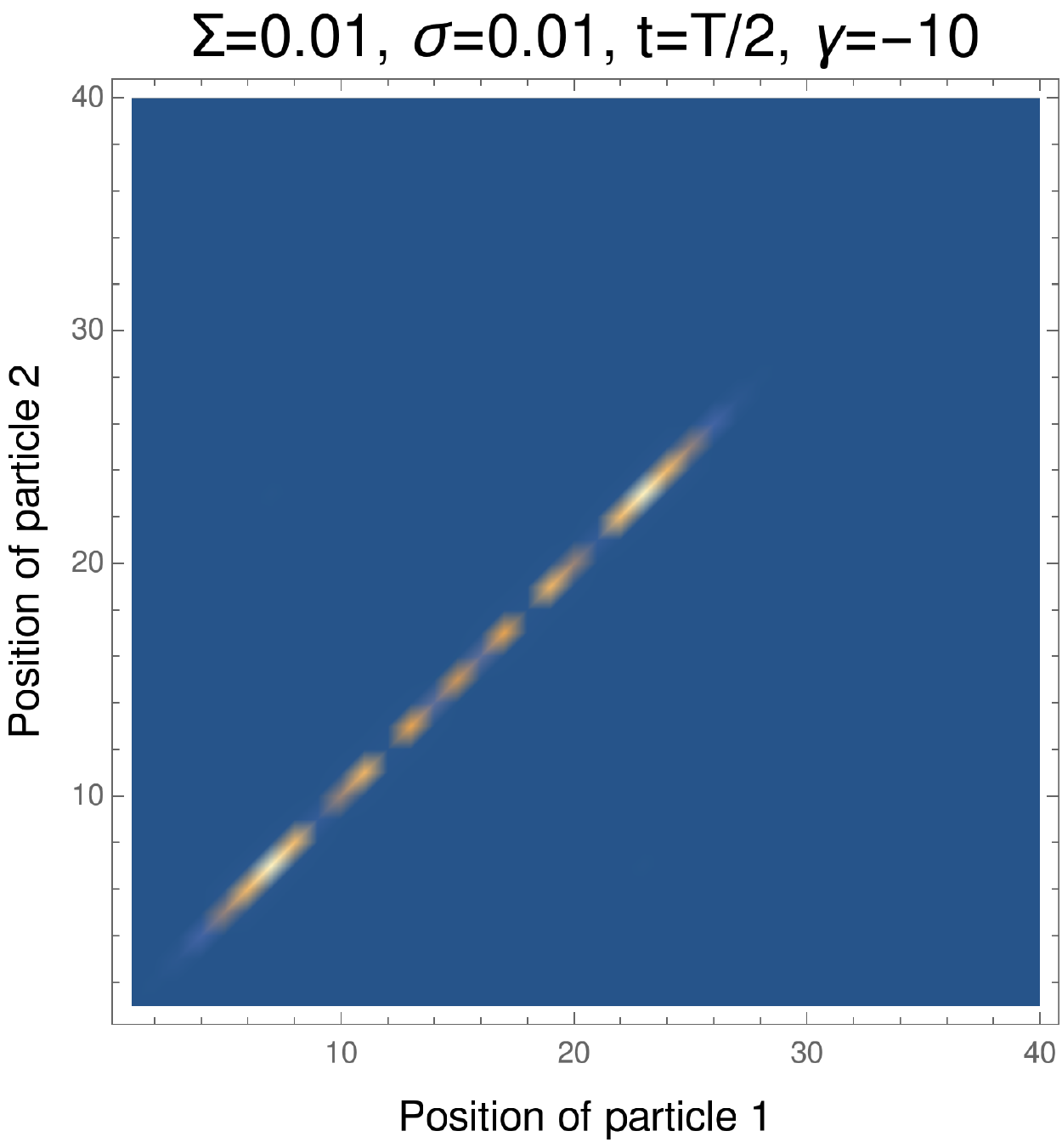}~\includegraphics[scale=0.22]{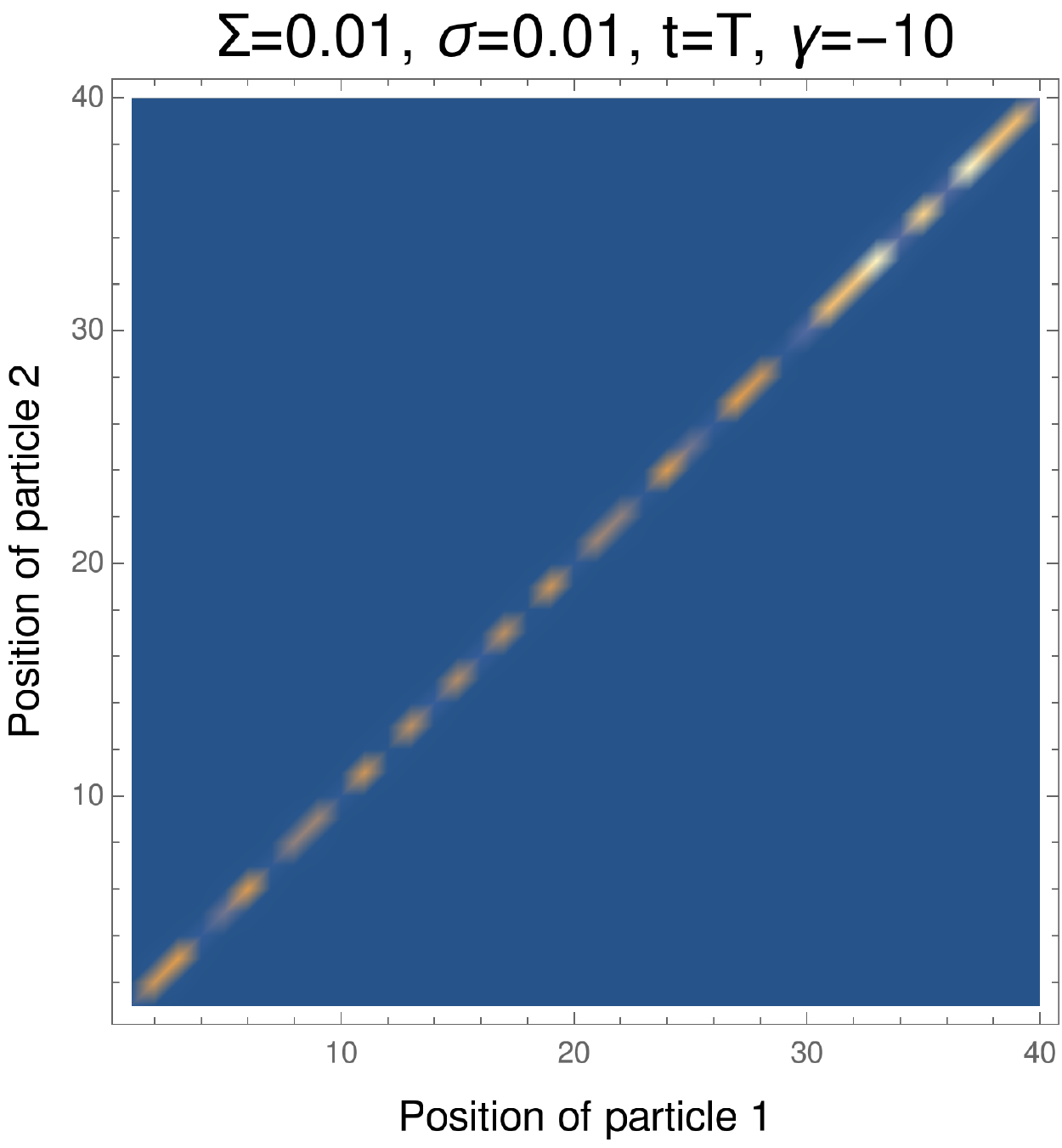}
\caption{Two-particle probability density plots showing the evolution in the MZI-like setup for $\varphi = 0$. The first row corresponds to entangled initial conditions, the second one to separable initial conditions, and the third one to the evolution with an interaction ($\gamma = -10$) between the particles. $T=11$ for the first two cases and $T=50$ for the last one. \label{fig5}}
\end{figure}

(II) Next, we evolve the system for time $T/2$ according to the unitary operator $U=e^{-i H_{free} T/2}$. The reason for introducing $T/2$ will become clear in a moment. We aim to realize a $50/50$ splitting to neighbouring cells $j=0$ and $j=2$, however this cannot be done with the perfect efficiency. This is because the perfect 50/50 splitting should be a periodic operation. The spectrum of $H_{free}$ is  
\begin{equation}
E(k_1,k_2)=-2\cos\left(\frac{2\pi}{d}k_1\right)-2\cos\left(\frac{2\pi}{d}k_2\right),
\end{equation}
where $k_1,k_2 = 0,1,\ldots,d-1$ correspond to the momenta of the first and the second particle. The ratios of the above eigenvalues are in general irrational, therefore the corresponding unitary operator $e^{-i H_{free} t}$ is quasi-periodic. Hence, we look for $T/2$ such that $e^{-i H_{free} 2 T} \approx \openone$. More precisely, due to the fact that the evolution takes place in the finite space with periodic boundary conditions, the initial wave packet should spread and after some time come back to the initial position. This recurrence need not be perfect, however we choose the time of the first recurrence as $2T$. In addition, due to periodic boundary conditions, at time $T$ the wave-packet should localize in the opposite cell ($j=3$). Therefore, at time $T/2$ the wave packet should be between $j=1$ and $j=3$, i.e., in a superposition of being in cells $j=2$ and $j=0$ (see Fig.~\ref{fig5}, first row). 

(III) After splitting the wave-packet to cells $j=0$ and $j=2$, we apply the phase shift in the cell $j=2$ generated by the following potential
\begin{equation}
V=\sum_{x_1,x_2=d/2+1}^{3d/4} (a_{x_1}^{\dagger}a_{x_1} + b_{x_2}^{\dagger}b_{x_2}).
\end{equation}
The corresponding unitary operator is of the form $U_{\varphi} = e^{i V \varphi}$, where $-\varphi=t$ is the time during which we apply the potential. 

(IV) We apply once again the free evolution operator  $U=e^{-i H_{free} T/2}$. Because such evolution generated splitting to neighbouring cells $j=0$ and $j=2$, now it must generate splitting to cells $j=1$ and $j=3$. This is because of translational symmetry and periodic boundary conditions. (V) Finally, we measure the operator $D_3$. 
 

\begin{figure}
\includegraphics[scale=0.22]{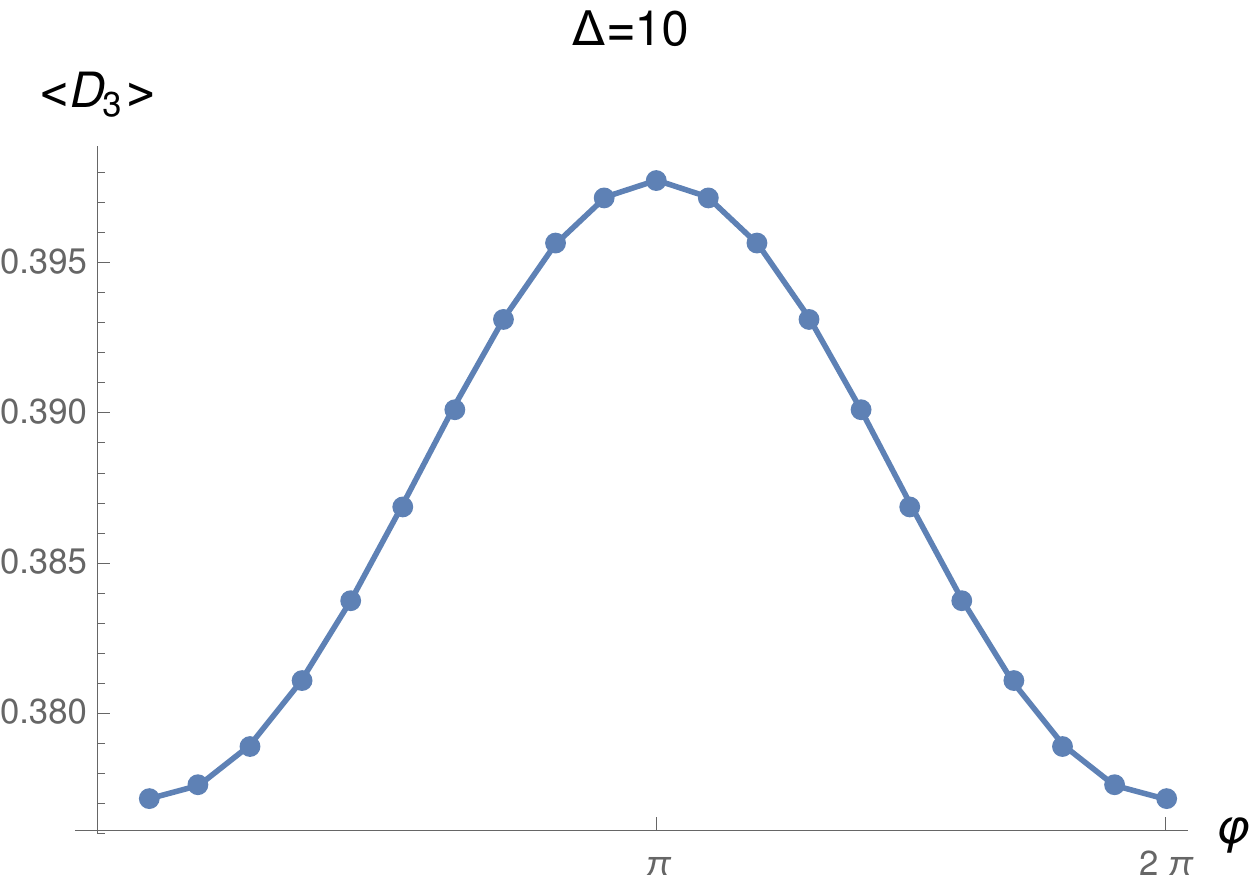}~\includegraphics[scale=0.22]{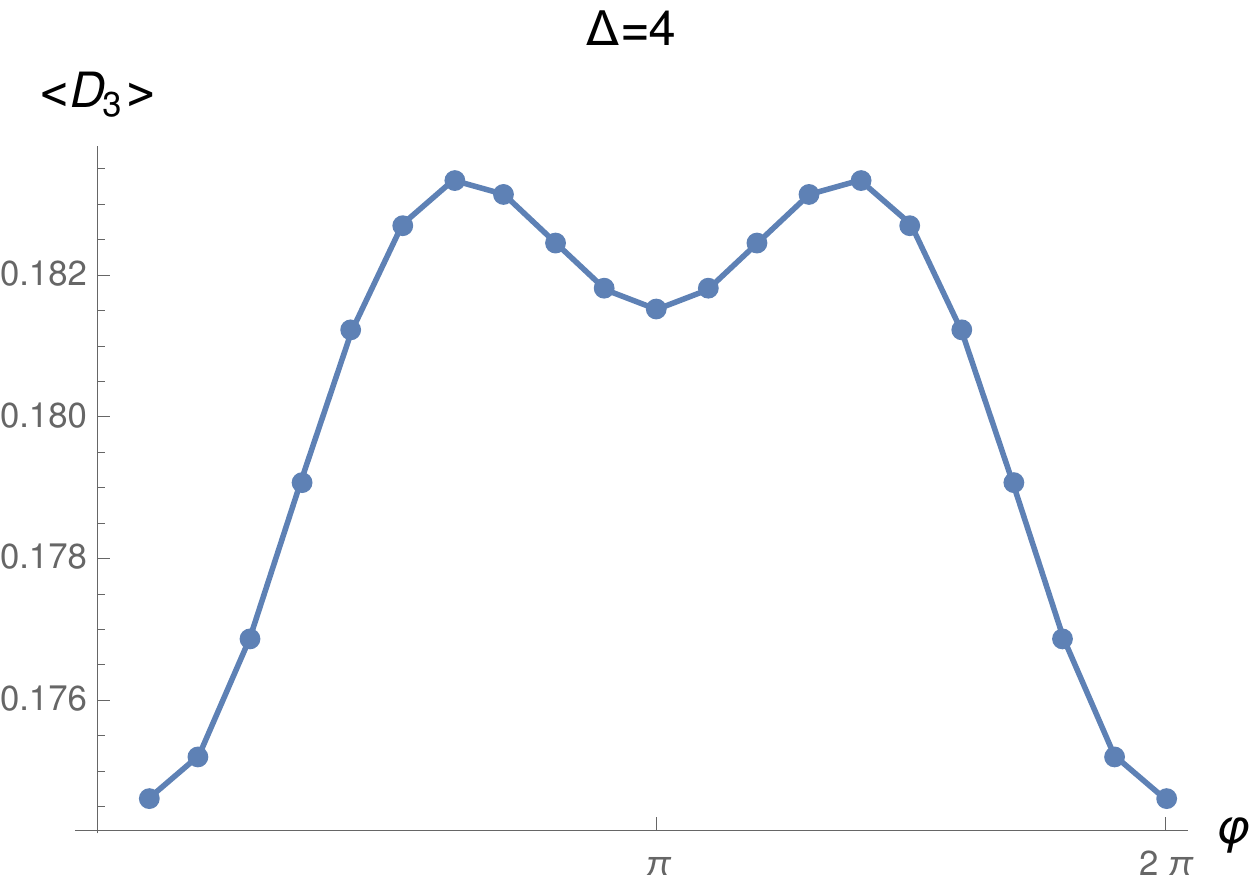}~\includegraphics[scale=0.22]{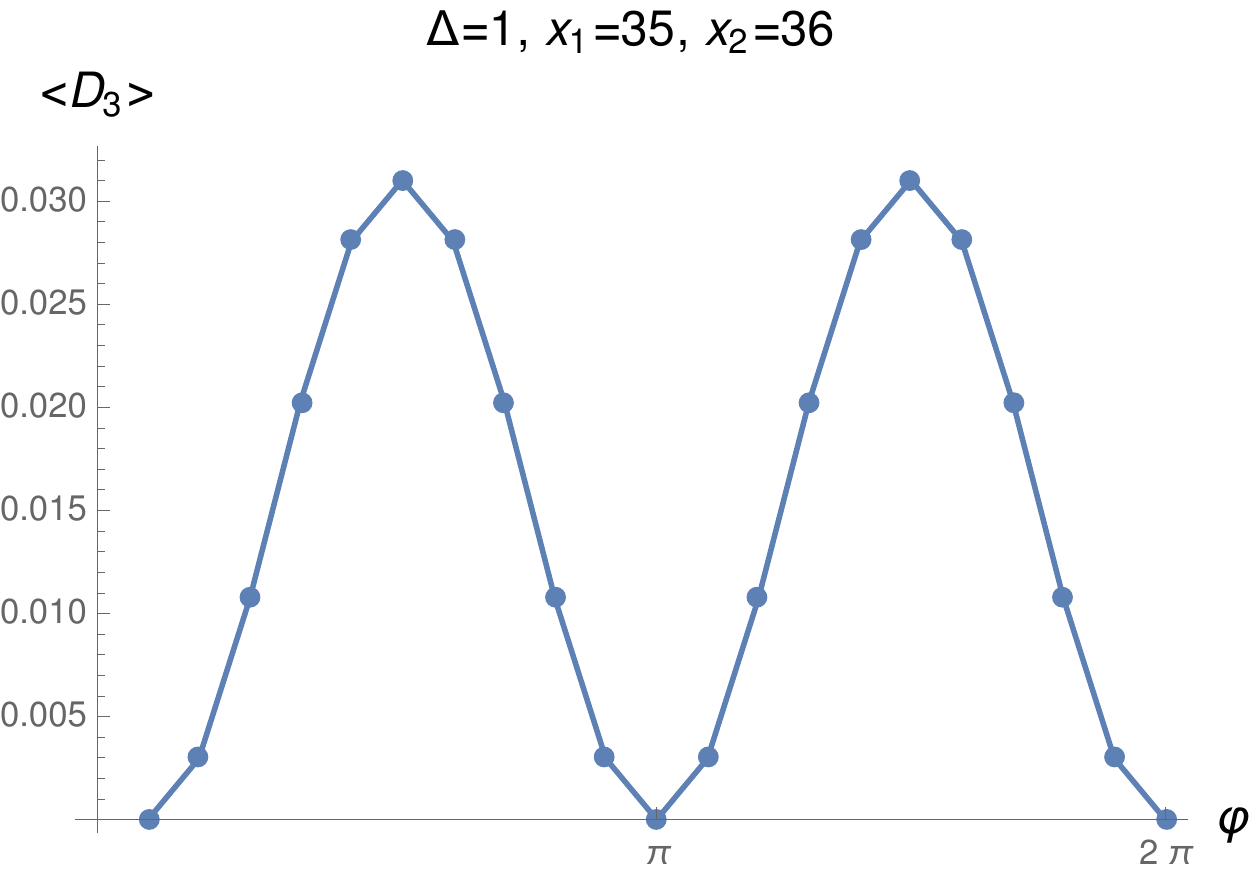}
\includegraphics[scale=0.22]{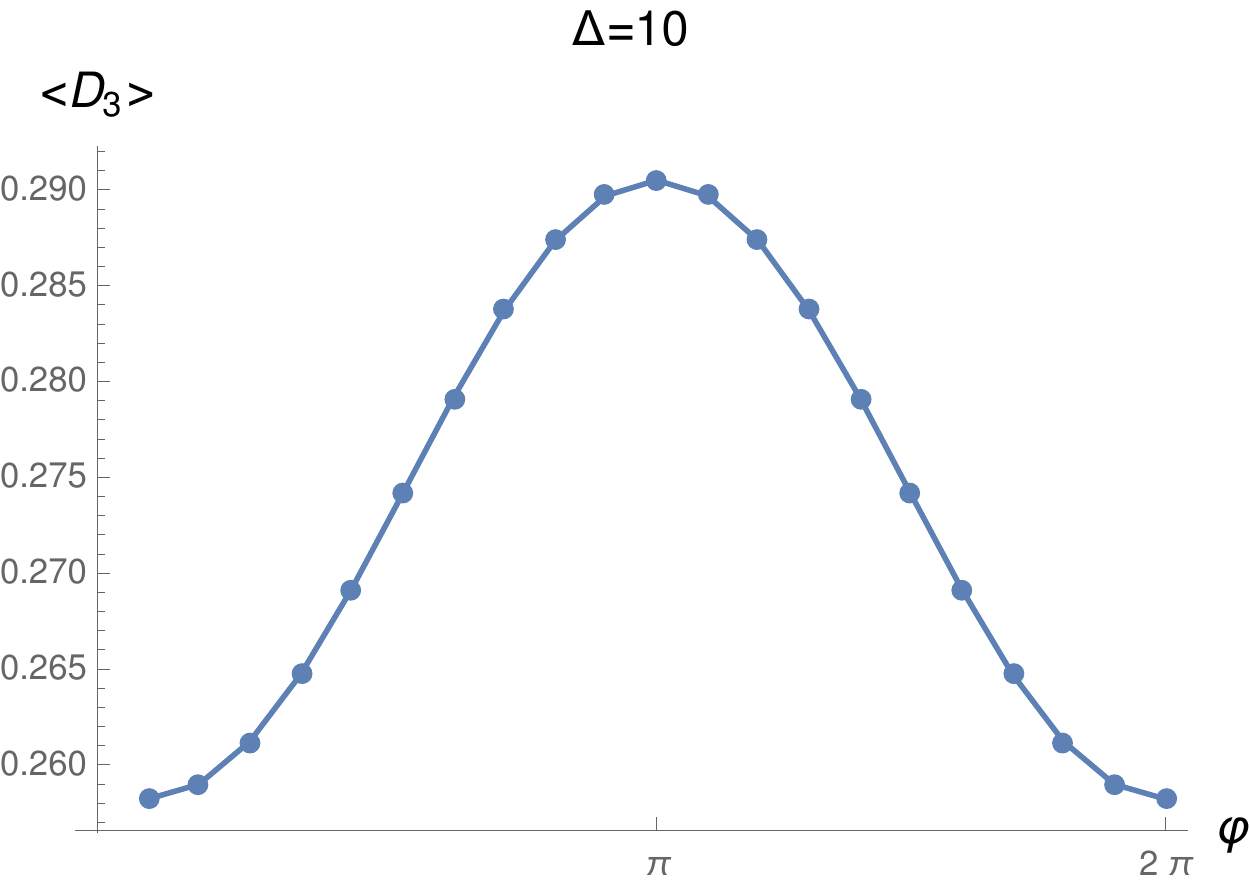}~\includegraphics[scale=0.22]{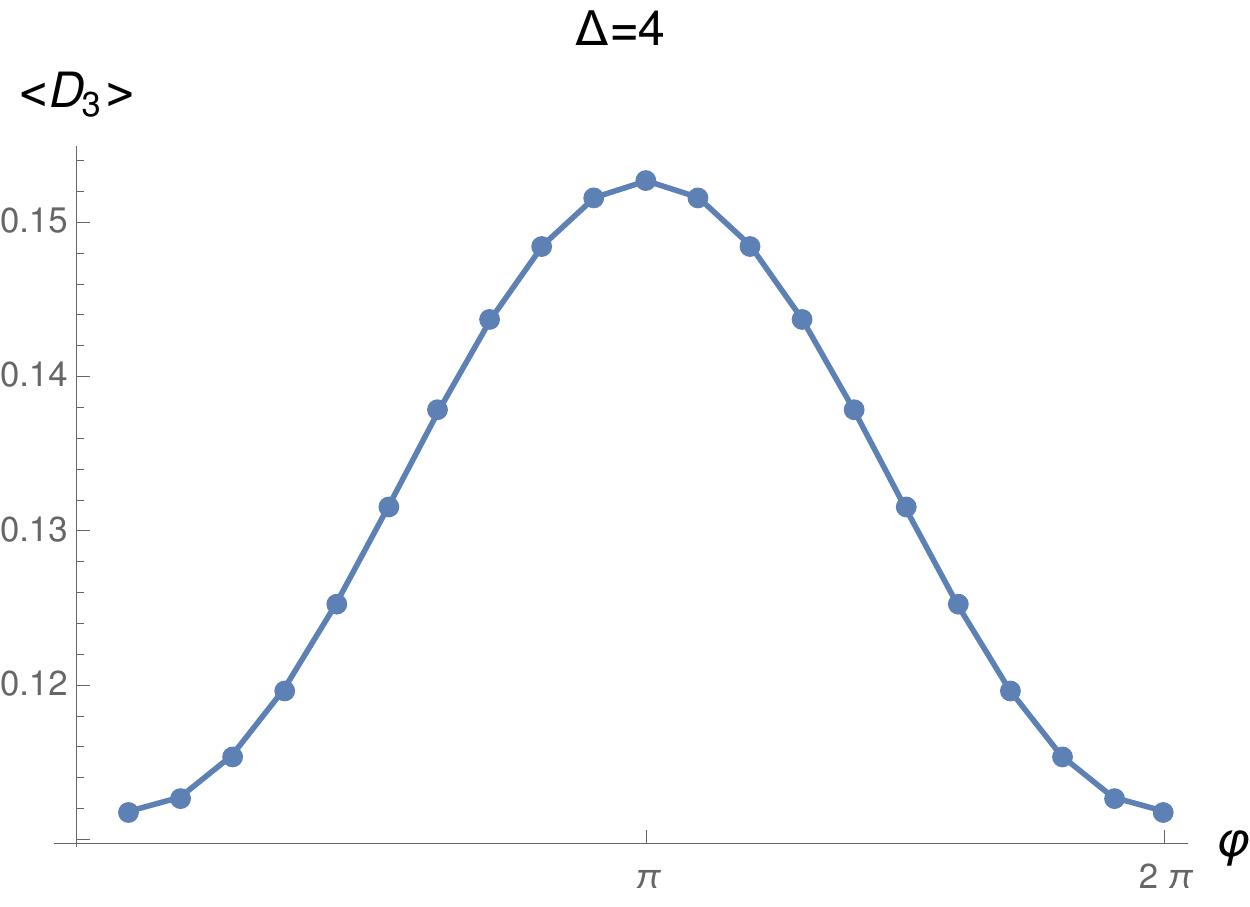}~\includegraphics[scale=0.22]{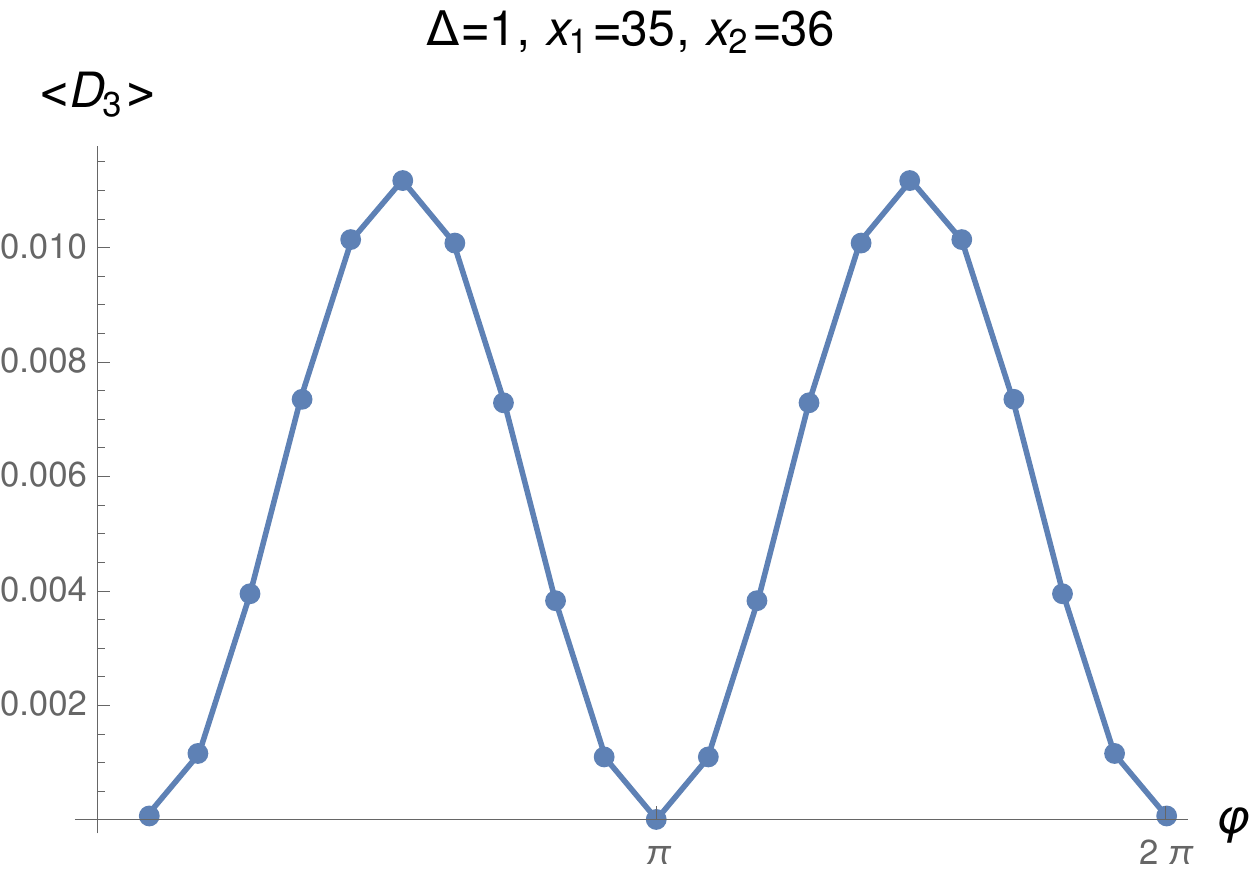}
\includegraphics[scale=0.22]{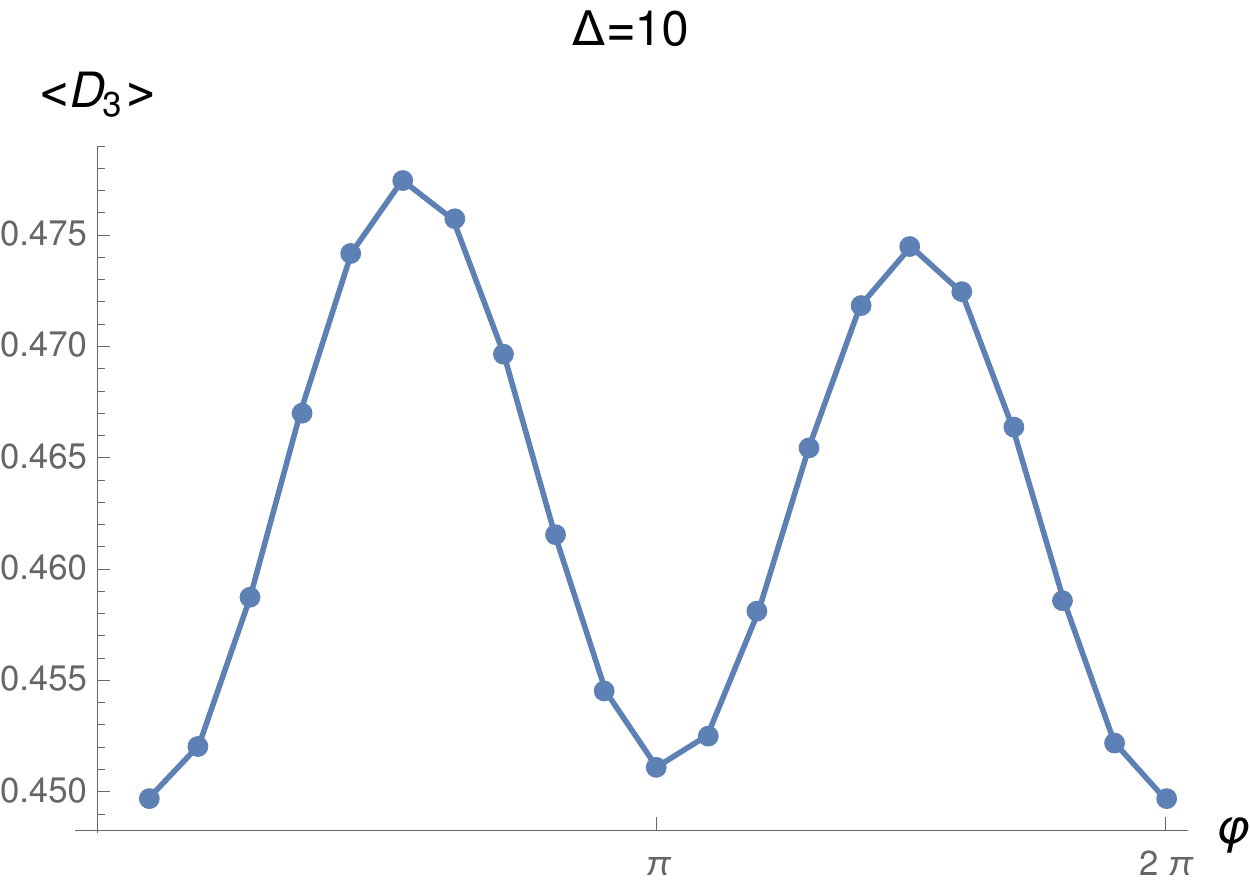}~\includegraphics[scale=0.22]{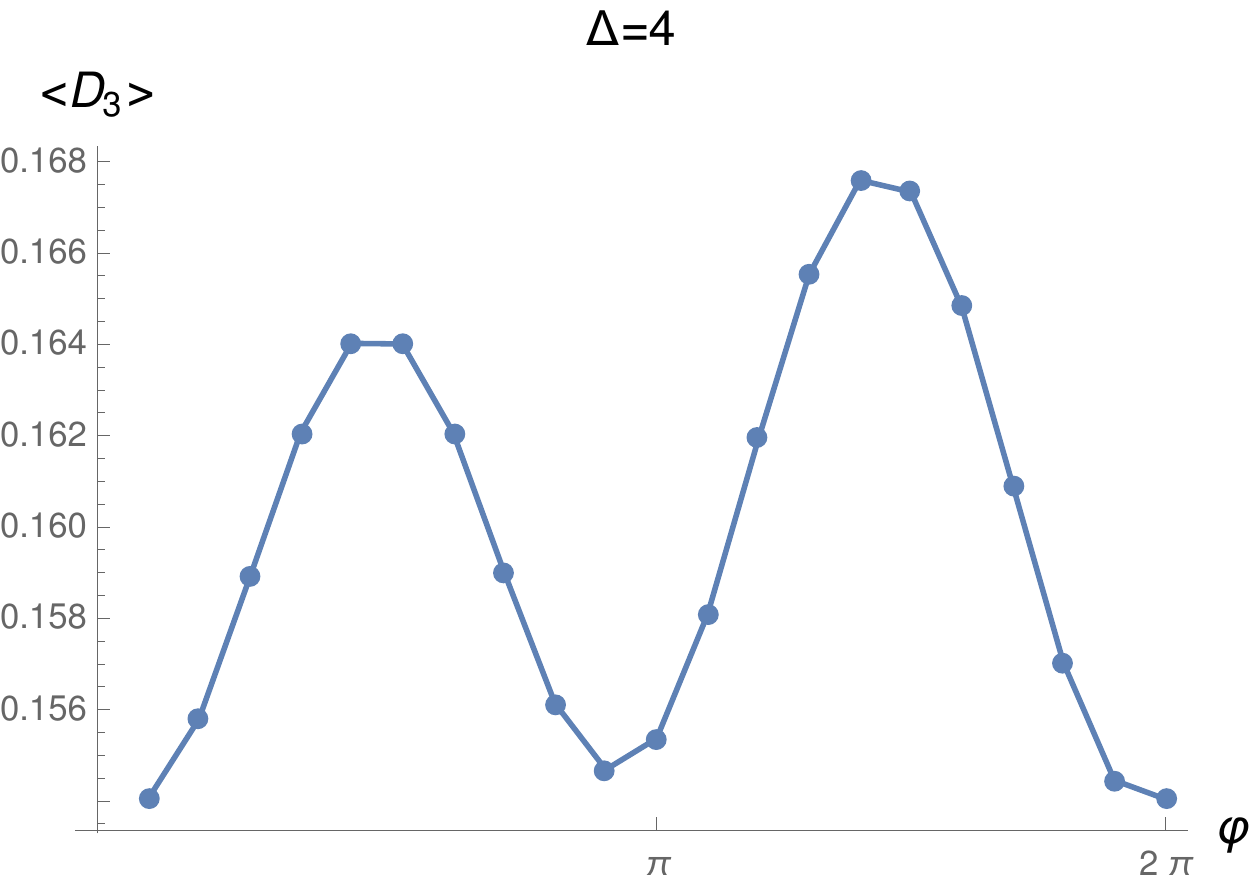}~\includegraphics[scale=0.22]{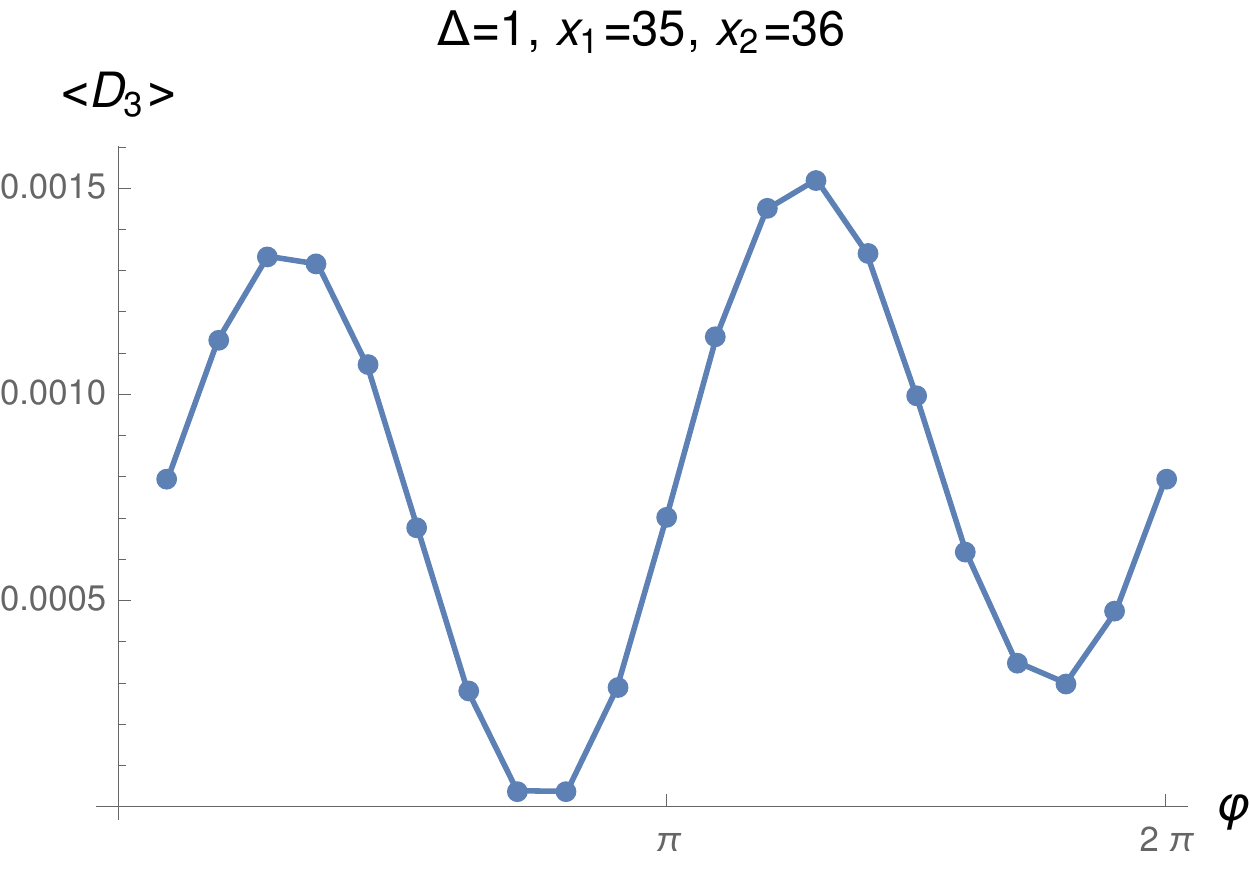}
\caption{Average values $\langle D_3 \rangle$ as functions of $\varphi$. The first row corresponds to entangled initial conditions, the second one to separable initial conditions, and the third one to the evolution with an interaction ($\gamma = -10$) between the particles. The graphs in each row represent different coarse graining of space which corresponds to different resolutions of the detector $D_3$. The coarse graining effect is especially important in the first two cases with no interaction. \label{fig6}}
\end{figure}


The dynamics of the above system was simulated numerically for $d=40$. We primarily focused on two situations -- entangled ($\Sigma = 0.01$ and $\sigma = 2$) and separable ($\Sigma = 0.01$ and $\sigma = 0.01$) initial conditions. For these two cases $T=11$. In addition, to complement our presentation, we also considered a dynamics of interacting particles (with the initial state being the same as in the separable case). This time, the Hamiltonian $H_{free}$ was supplemented with the on site interaction term of the form
\begin{equation}\label{int}
H_{int} = \gamma \sum_{x=1}^d a^{\dagger}_x a_x b^{\dagger}_x b_x,
\end{equation}
where $\gamma$ is the interaction strength ($\gamma > 0$ repulsion and $\gamma < 0$ attraction). We chose $\gamma=-10$ for which $T=50$. The evolution of the corresponding wave-packet for $\varphi=0$ is presented in Fig. \ref{fig5}.

The interference pattern produced at the detector $D_3$, i.e., the value of $\langle D_3 \rangle$ as a function of $\varphi$, is presented in Fig. \ref{fig6}. We considered three different detection strategies to show how the ability to see the internal structure of the composite particle affects the interference pattern. In the first case we choose to observe the whole cell ($\Delta = 10$), which means that the detector clicks if both particles are in the coarse grained cell $j=3$ ($31 \leq x_1,x_2 \leq 40$). For the second case we choose  $\Delta = 4$ for which the detector clicks if both particles are in the region $34 \leq x_1,x_2 \leq 37$. Finally, for the third case we choose $\Delta = 1$ with slightly shifted positions, i.e., the detector clicks if the first particle is at position $x_1=35$ and the second at position $x_2=36$.

For $\Delta=10$ only the interaction produces a doubled interference pattern, i.e., a pattern with a half-period. For $\Delta=4$ the situation changes and one starts to observe some doubling in case of entangled initial conditions. This is because the detection is limited to a region whose size is comparable to the size of the composite particle. Therefore, the detector starts to observe the internal structure of the composite particle. Finally, for $\Delta = 1$ we see doubling in all three cases, which might seem surprising for separable initial conditions. However, note that in case of separable states $\langle a_{x_1}^{\dagger}a_{x_1}b_{x_2}^{\dagger}b_{x_2} \rangle = \langle a_{x_1}^{\dagger}a_{x_1} \rangle \langle b_{x_2}^{\dagger}b_{x_2} \rangle$. In particular, if the probability that the first particle is at $x_1=35$ is a periodic function with the period $2\pi$ and the probability that the second one is at $x_2=36$ is another periodic function with period $2\pi$, it can happen that their product is a periodic function with period $\pi$. Such a possibility was already discussed above in case of the standard MZI. Therefore, the doubling in this case is due to a specific measurement, rather than the entanglement, and one cannot say that it is a result of an interference of a composite particle with itself. 

To conclude, we see that the only possibility to observe the de Broglie wavelength of a composite particle, without analysing its internal structure, was when the particles were interacting. The doubling of the interference pattern can be observed in non-interacting systems once we start to analyse the relations between the two elementary constituents, however such a phenomenon cannot be called a single-particle one. We will come back to this issue and we will show that this is a generic feature that does not depend on the form of an entangled wave-packet. 


\subsection{Bloch oscillations}

Next, we are going to study another effect that is known to occur in systems in which particles move on a lattice. We set $H=H_{free} + V$, where 
\begin{equation}
V=\eta\sum_{x=1}^d x\left(a_{x}^{\dagger}a_{x}+b_{x}^{\dagger}b_{x}\right)\label{BOH}
\end{equation}
is a potential which imitates constant force (due to constant electric field, etc.). The parameter $\eta$ determines the magnitude of the force.

The above Hamiltonian generates an interesting evolution. Because of discreteness of space the momentum is confined to the first Brillouin zone $\hbar k\in [-\frac{\hbar\pi}{a},\frac{\hbar\pi}{a})$, where $a$ is the lattice constant. We set $a=1$, therefore in our case $k\in [-\pi,\pi)$. Now, let us study the action of the unitary operator $U=e^{-i t \eta \sum_x x a_x^{\dagger}a_x}$ on single-particle momentum eigenstates $|k\rangle$. Since $e^{ix\delta}|k\rangle = |k+\delta\rangle$, therefore  $U|k\rangle = |k - t\eta \rangle$. However, because momentum is in the first Brillouin zone, we get $k - t\eta \equiv k - t\eta + m2\pi$. This has important implications. We observe that particle starts to oscillate. It accelerates in one direction and, after reaching the border of the Brillouin zone, it suddenly changes the direction of its movement. The force causing acceleration is still acting, therefore the particle, which moves in the other direction, slows down and stops. Finally, it starts to accelerate in the original direction once more. This periodic oscillatory motion is known as Bloch oscillations. The period of the oscillations is $T_{BO}=\frac{2\pi}{\eta}$.  

It is known that in systems of interacting particles it is possible to observe Bloch oscillations with fractional periods $\frac{2\pi}{N \eta}$, where $N$ is the number of particles \cite{claro03,dias07,khomeriki10,longhi11,ahlbrecht12,preiss15}. This is a composite effect analogous to the observation of the collective de Broglie wavelength in the MZI. However, it is not clear what kind of composite effects persist in Bloch oscillations if the constituent
particles do not interact but are entangled. Although Bloch oscillations of multiple entangled non-interacting photons were considered before \cite{bromberg10,lebugle15}, the authors of these works focused on relative position between the photons, which in our case corresponds to the internal structure of the composite particle. They found, that one can observe periodic transitions between bunching and antibunching with a period being a fraction of $T_{BO}$. Here, we are going to study numerically the evolution generated by the above Hamiltonian on a state (\ref{gauss2}) and are going to use measurements (\ref{detector}) to see if it is possible to observe any composite effects without studying the internal structure of the system.


\begin{figure}
\includegraphics[scale=0.22]{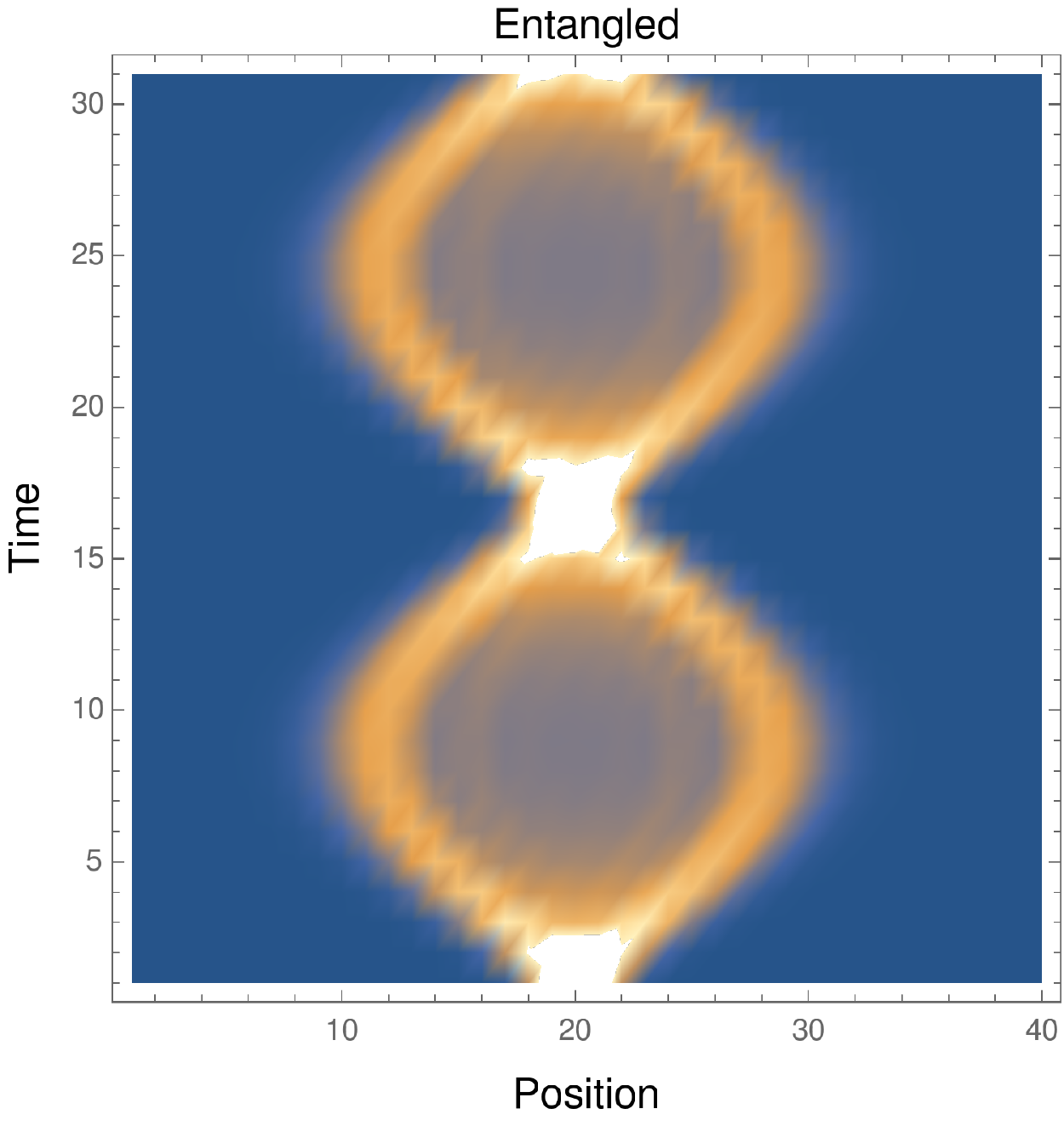}~\includegraphics[scale=0.22]{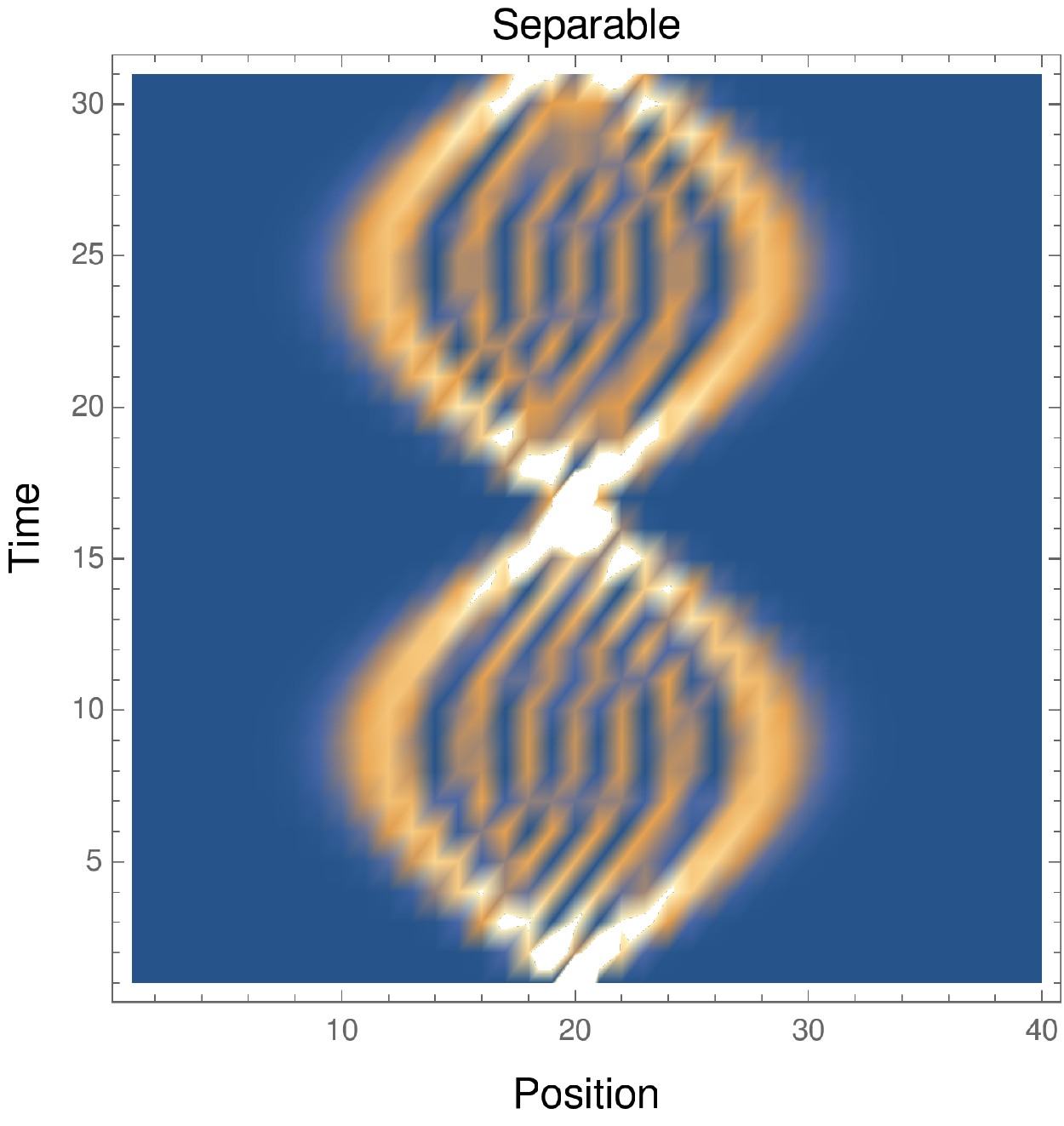}~\includegraphics[scale=0.22]{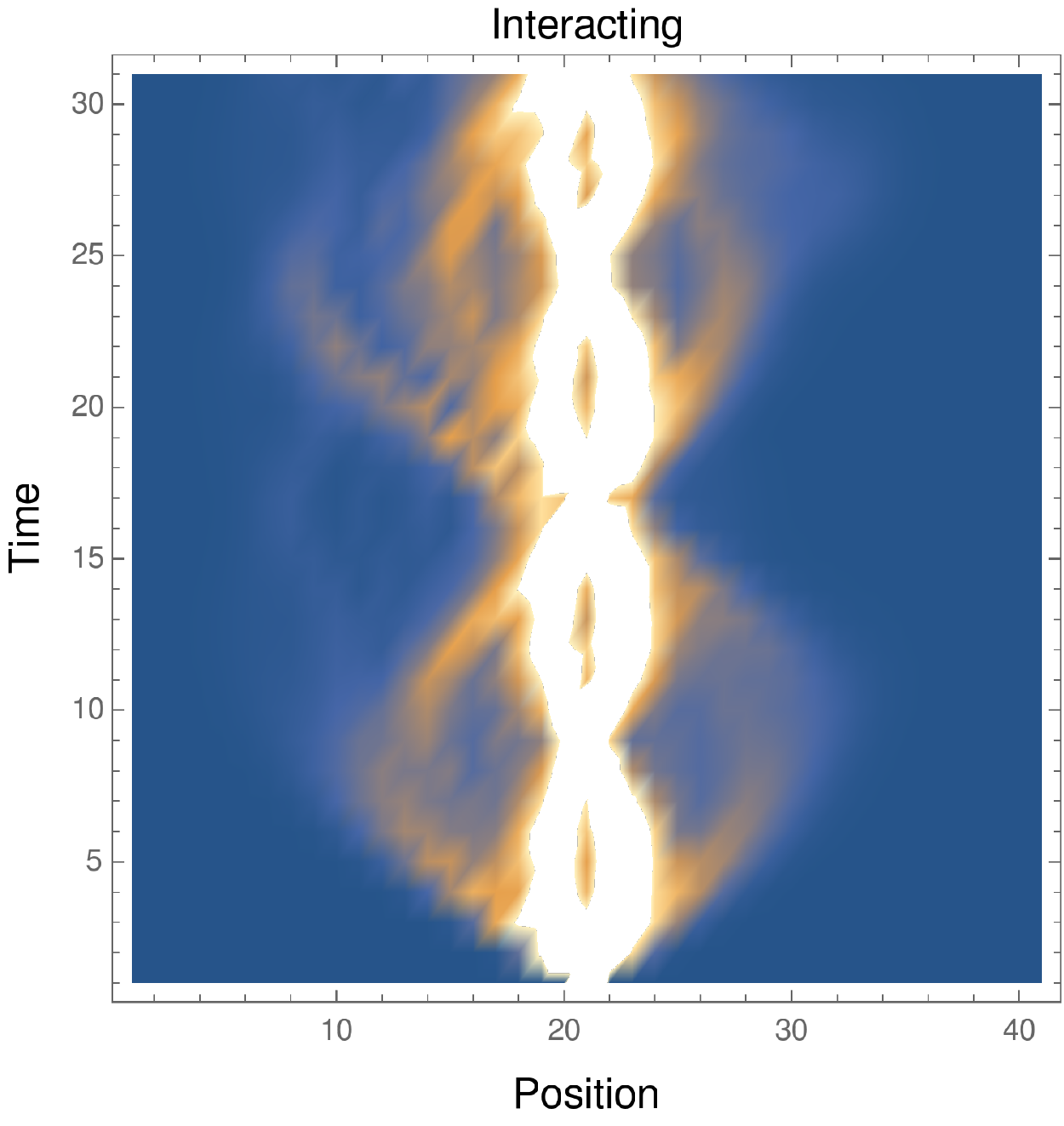}
\caption{Time evolution of the particle density plots $\langle a_x^{\dagger}a_x \rangle + \langle b_x^{\dagger}b_x \rangle$ for $\eta = 0.4$, which gives $T_{BO}\approx 15.7$. Left: non-interacting particles in the entangled state (\ref{gauss2}) with $\sigma = 2$, $\Sigma = 0.01$ and initially centred around $x=20$. Center: non-interacting particles in the separable state (\ref{gauss2}) with $\sigma = 0.01$, $\Sigma = 0.01$ and initially centred around $x=20$. Right: interacting particles with $\gamma = -2.5$ and initial state $a^{\dagger}_{20} b^{\dagger}_{20} |0\rangle$. Bloch oscillations with a fractional period are only visible when interaction is turned on. \label{fig7}}
\end{figure}


\begin{figure}
\includegraphics[scale=0.22]{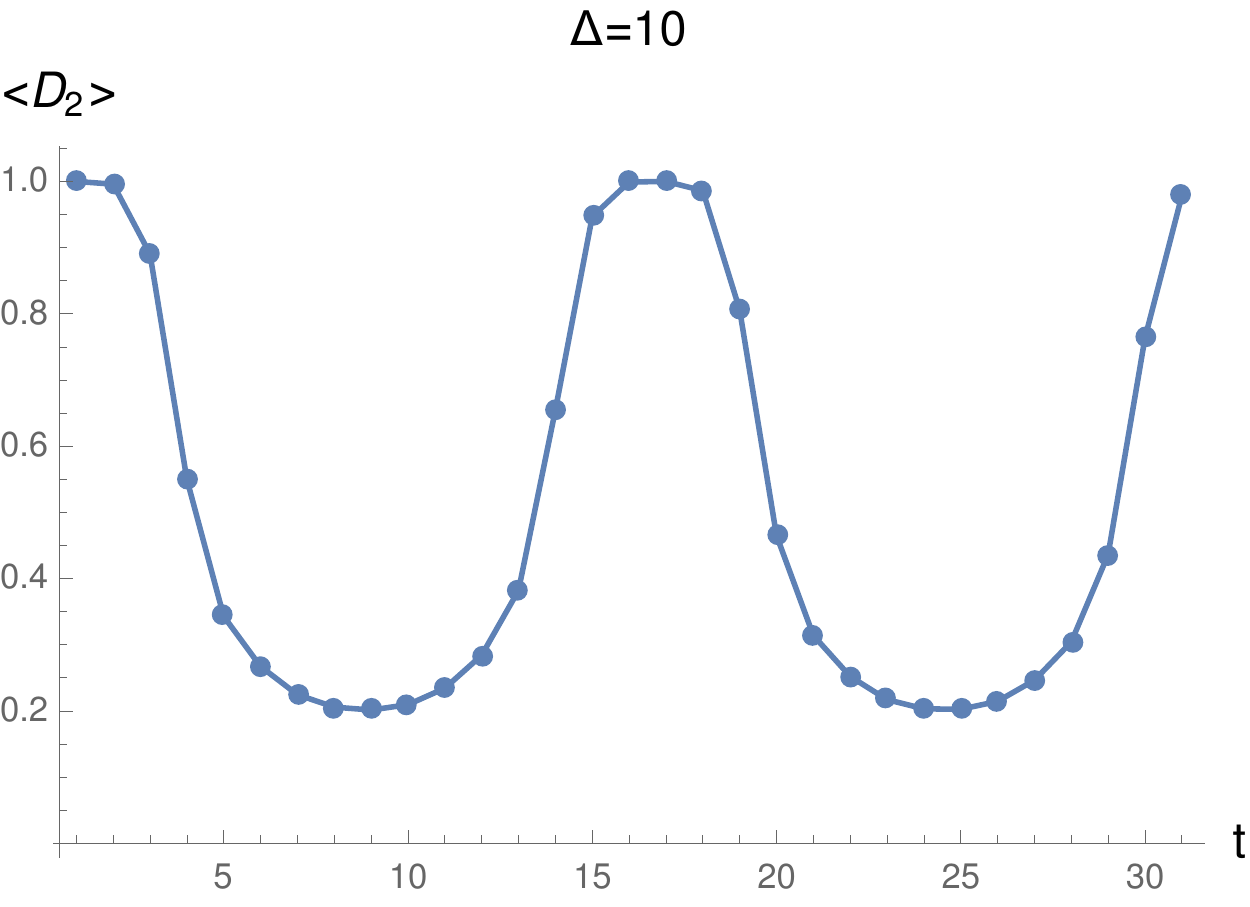}~\includegraphics[scale=0.22]{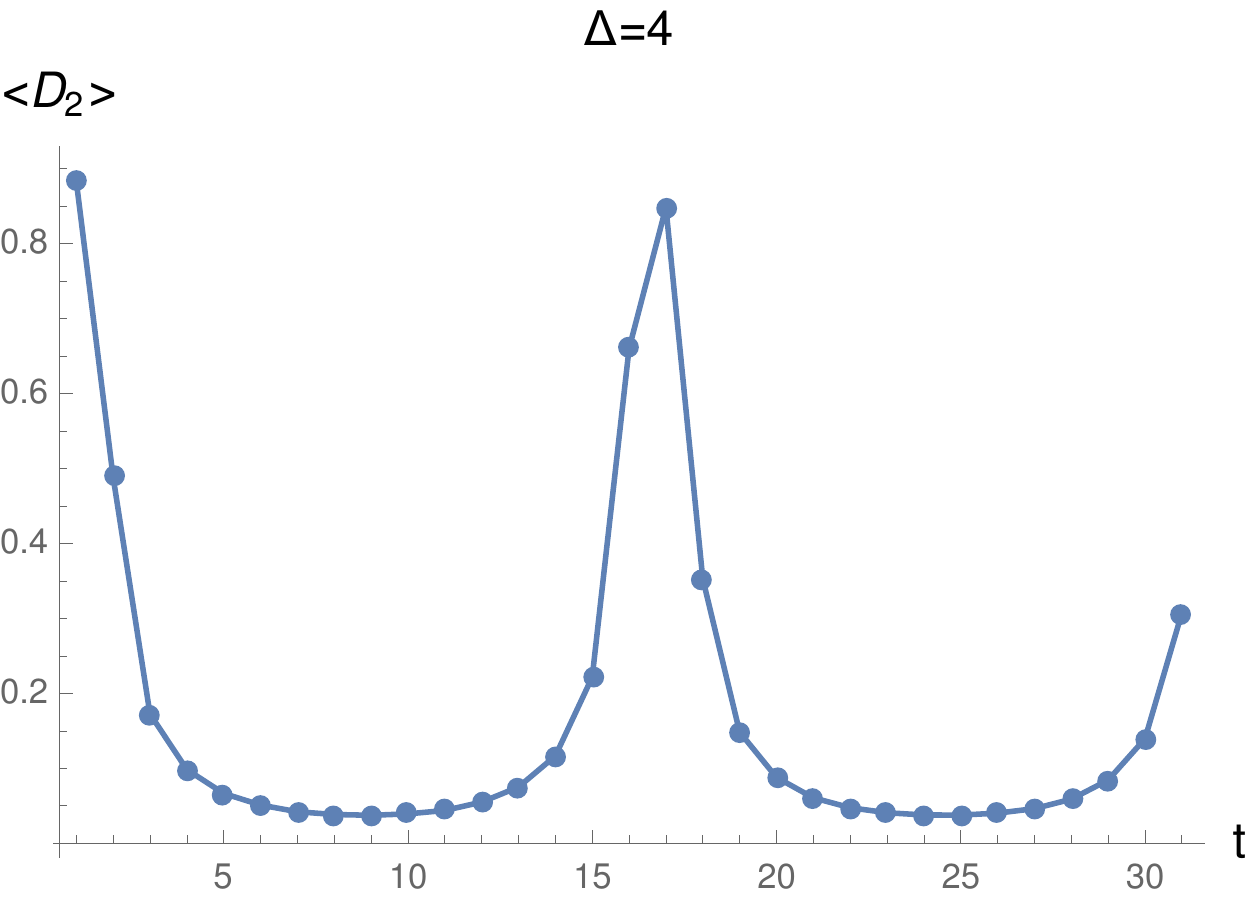}~\includegraphics[scale=0.22]{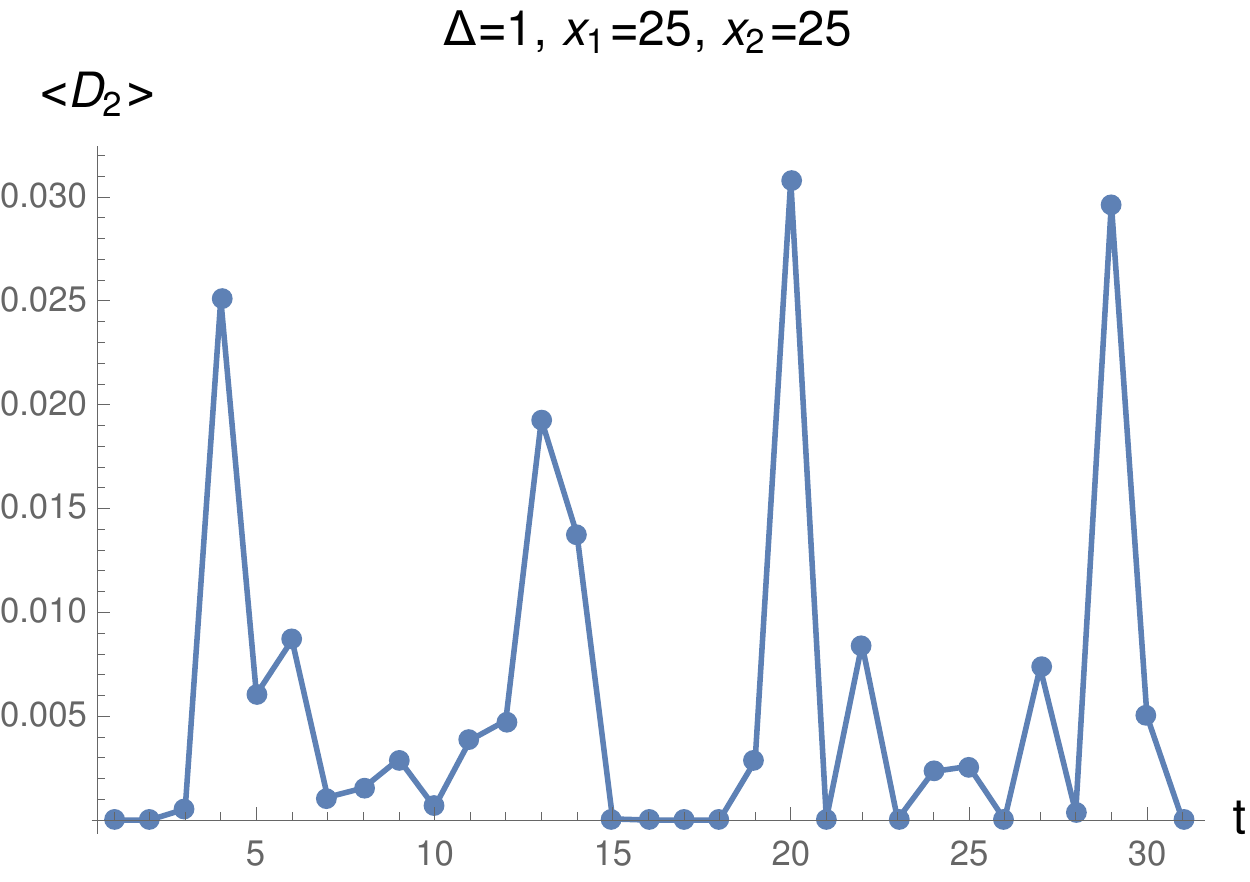}
\includegraphics[scale=0.22]{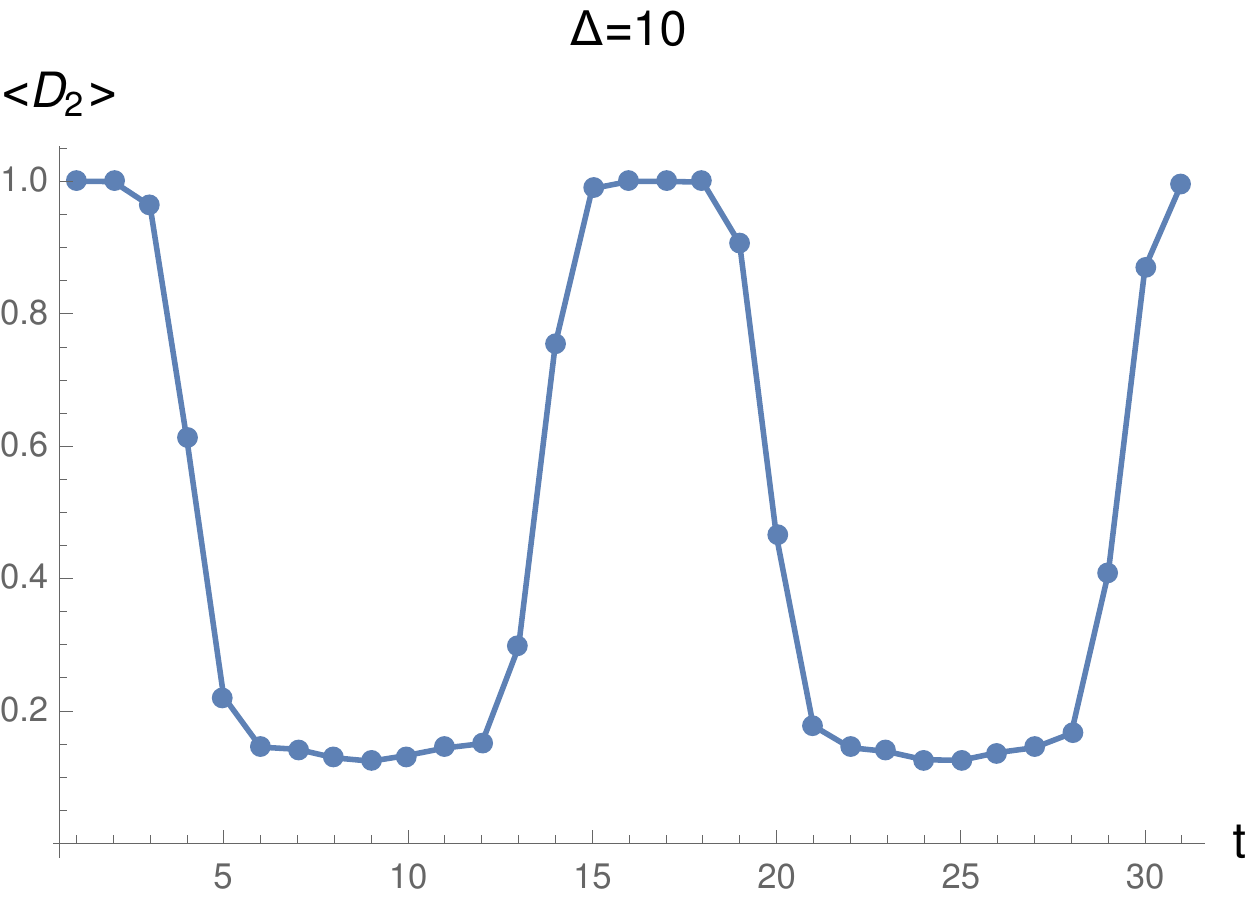}~\includegraphics[scale=0.22]{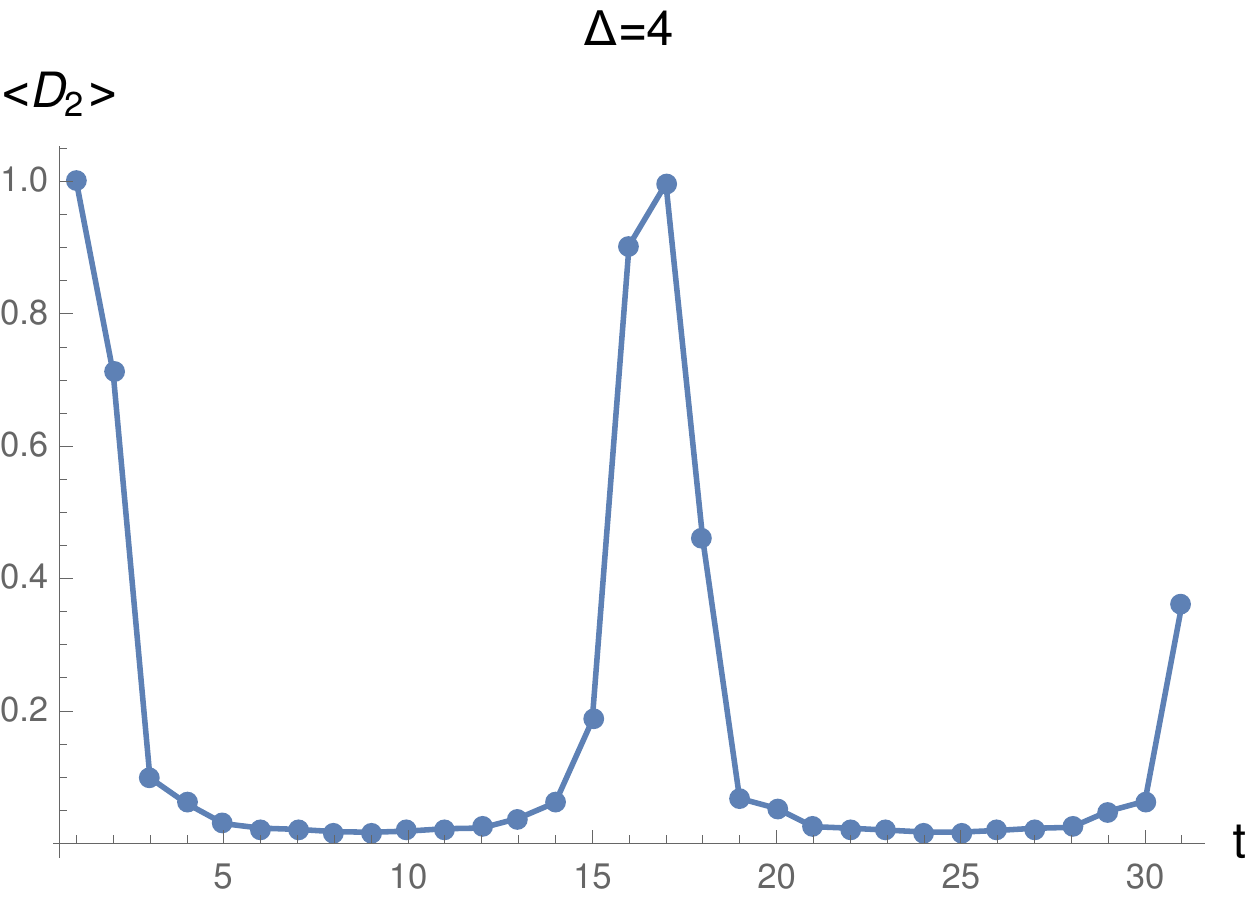}~\includegraphics[scale=0.22]{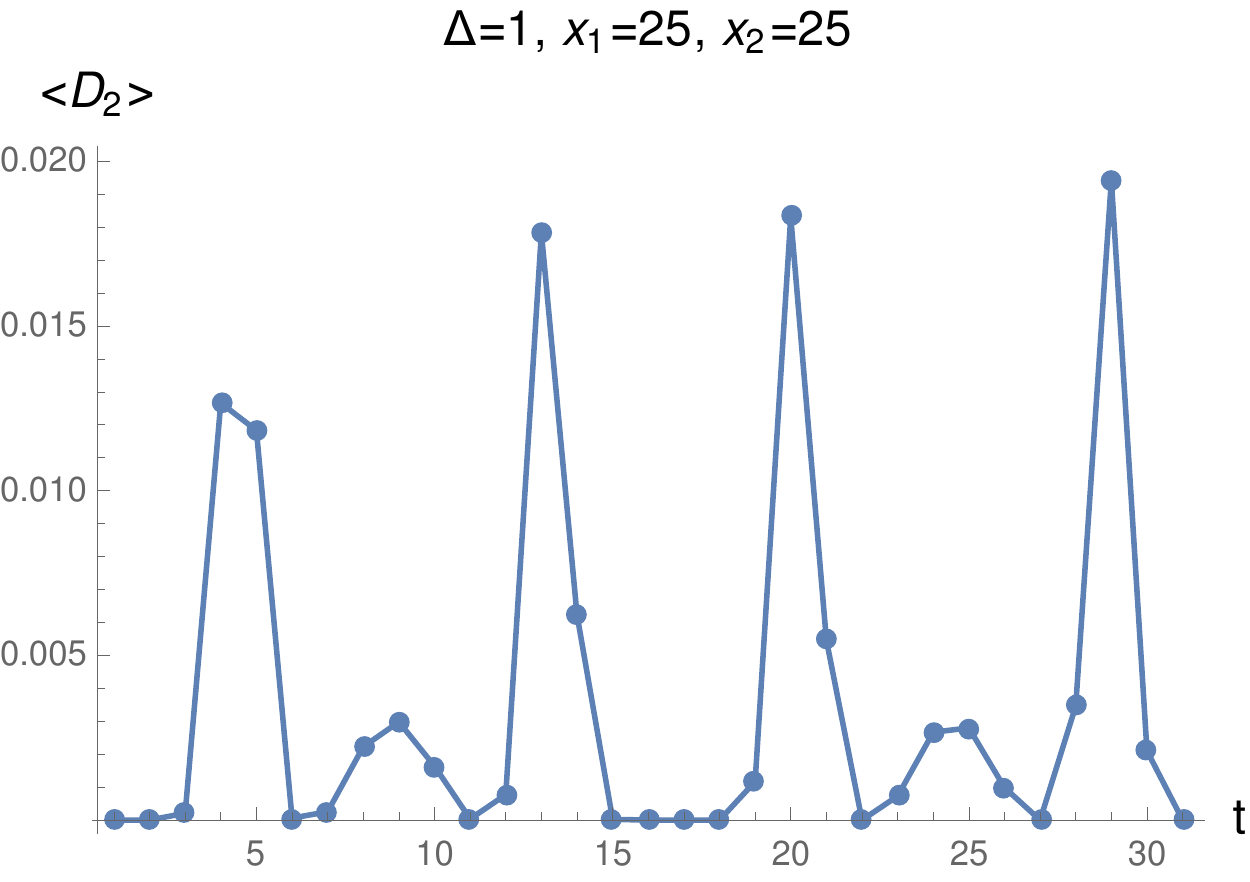}
\includegraphics[scale=0.22]{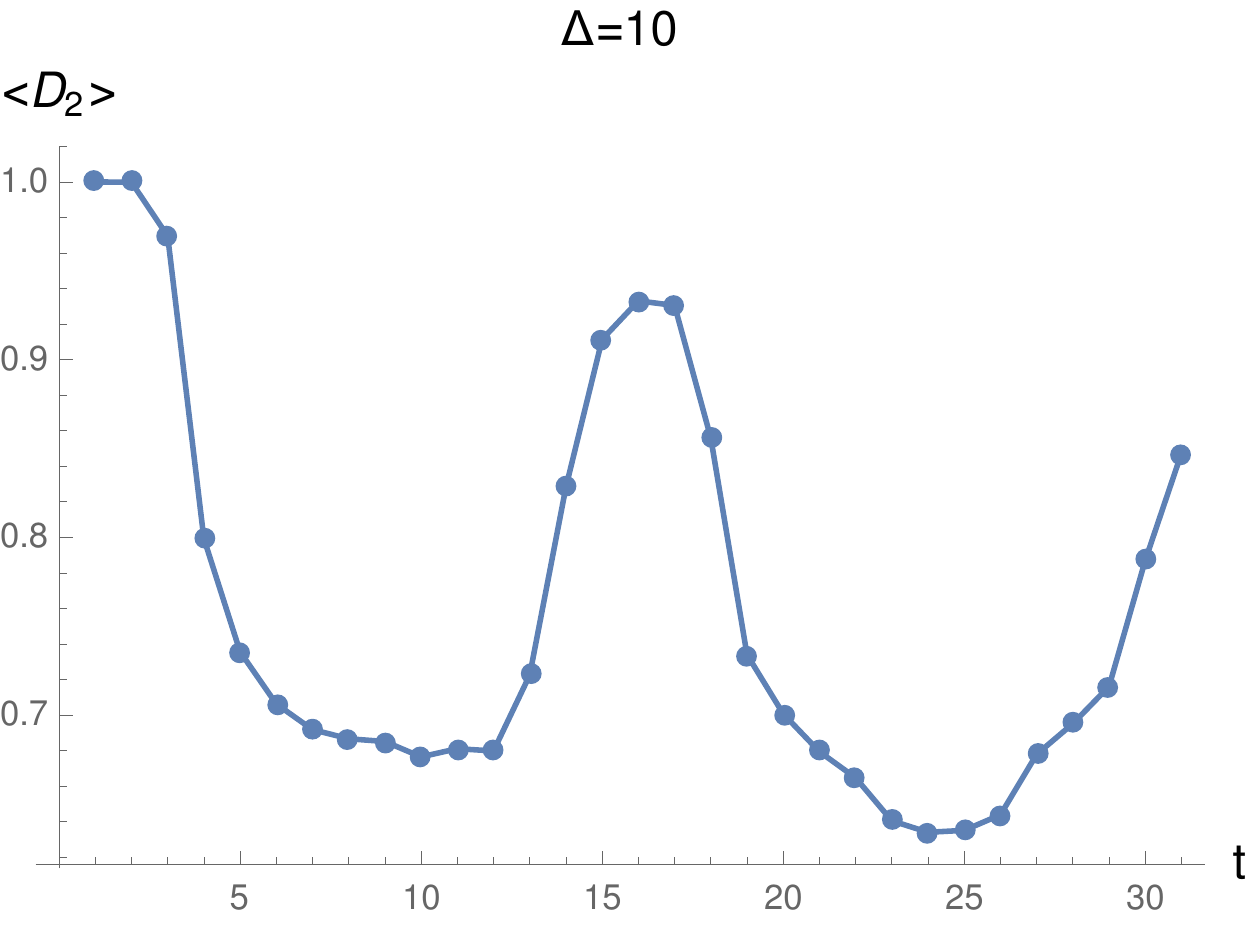}~\includegraphics[scale=0.22]{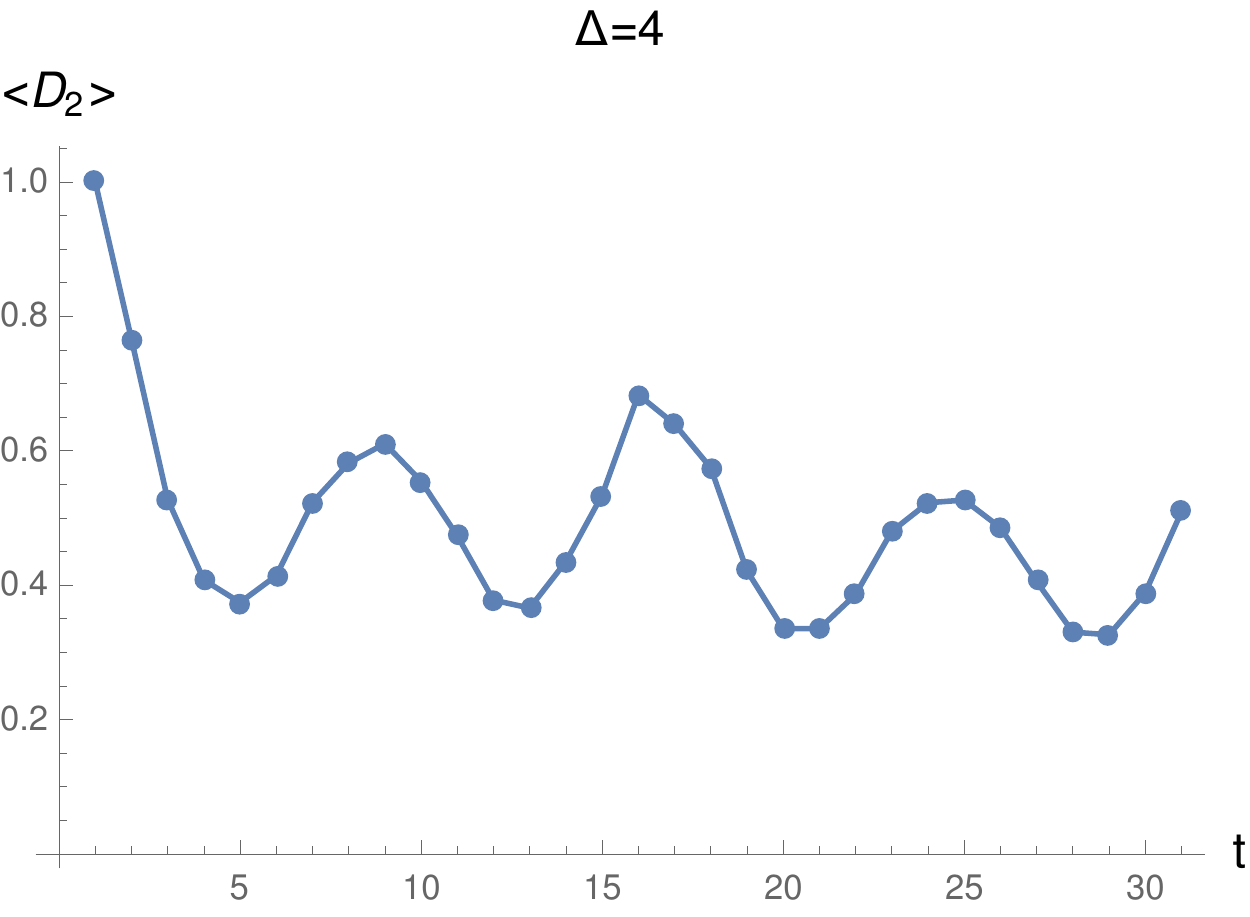}~\includegraphics[scale=0.22]{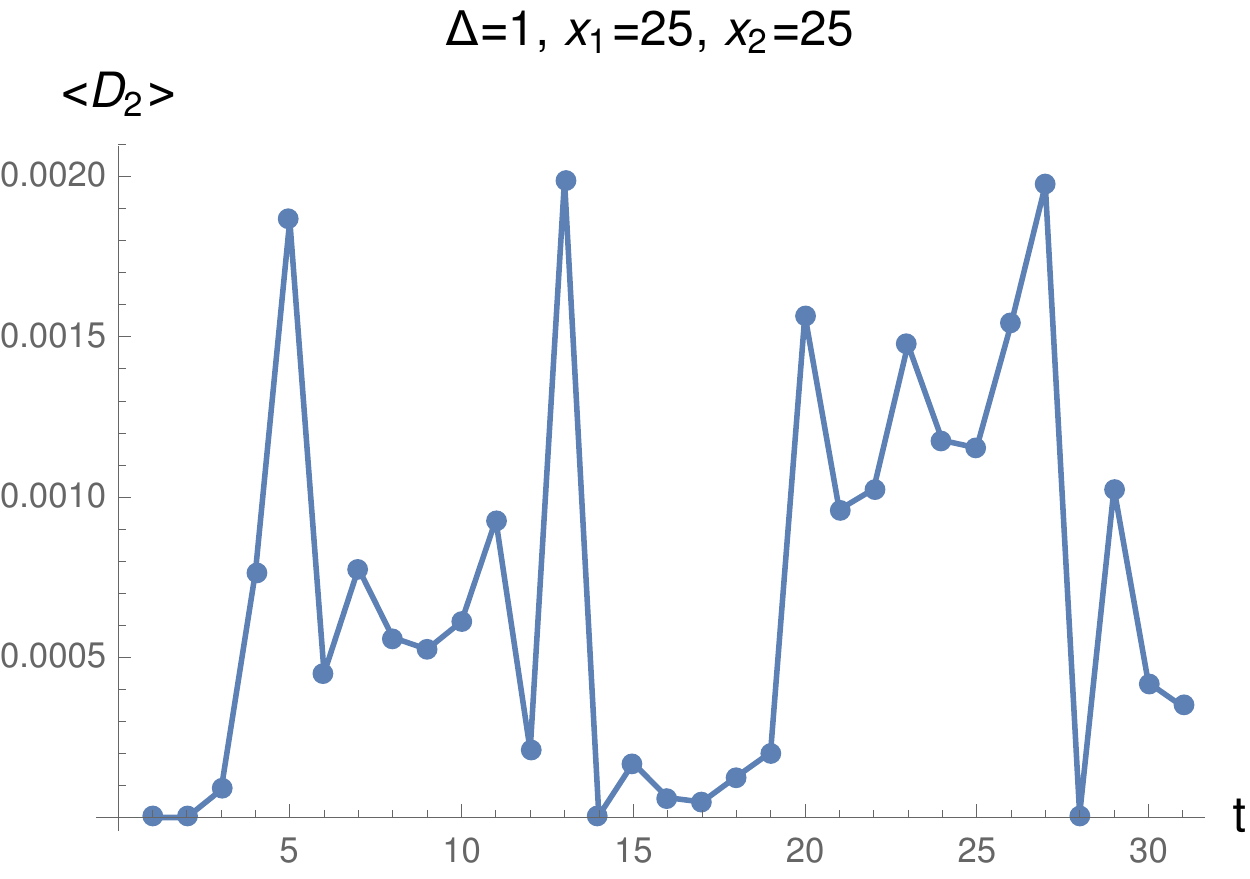}
\caption{Time evolution of the average value $\langle D_2 \rangle$. The first row corresponds to entangled initial conditions, the second one to separable initial conditions, and the third one to the evolution with an interaction ($\gamma = -2.5$) between the particles. The graphs in each row represent different coarse graining of space which corresponds to different resolutions of the detector $D_2$. The oscillations with a fractional period can be observed only when interaction is turned on, provided the observed region is smaller than the amplitude of oscillations (amplitude is larger than the resolution of the detector). It is possible to observe more complex oscillations if one can detect exact positions of both particles (the third collumn). \label{fig8}}
\end{figure}


As in the MZI case, we choose $d=40$ and $\Delta=10$. The space is divided into four coarse grained cells $j=0,1,2,3$ corresponding to detection operators (\ref{detector}), but this time the respective cell centres are $x=0,10,20,30$. Moreover, we choose the magnitude of the external force to be $\eta=0.4$, which corresponds to $T_{BO}\approx 15.7$. We prepare the initial wave packet (\ref{gauss2}) centred around $x=20$ (cell $j=2$). As before, we consider three cases: entangled initial conditions ($\Sigma=0.01$, $\sigma=2$), separable initial conditions ($\Sigma=0.01$, $\sigma=0.01$), and separable initial conditions with attraction between the particles ($\gamma=-2.5$). 

In Fig. {\ref{fig7}} we plot the time evolution of the particle density $\langle a^{\dagger}_x a_x + b^{\dagger}_x b_x\rangle$. Bloch oscillations are clearly visible, however the oscillations with a fractional period occur only when particles interact. The only difference between separable and entangled initial conditions is the interference pattern inside the oscillating envelope, which can be attributed to the fact that for separable state each particle is in a coherent superposition of momentum states, whereas for entangled state each particle is in a mixture of such states. 

In Fig. {\ref{fig8}} we show time evolution of average values $\langle D_2 \rangle$ and how they are affected by the change of $\Delta$. Here, we see basically no difference between the entangled and the separable initial conditions (the first and the second row). For these two cases we see Bloch oscillations with a period $T_{BO}$, as long as exact positions of both particles cannot be determined. One can observe some $\Delta$-dependence in the width of peaks, because for large $\Delta$ the system needs more time do exit the region $j=2$. However, one can observe more complex oscillations if the exact positions of both particles are known (we considered positions $x_1=x_2=25$). Interestingly, the oscillations with a fractional period are visible if the particles interact and only for the values of $\Delta$ which are smaller than the amplitude of oscillations (here $\Delta = 4$). The amplitude of the oscillations with a fractional period is smaller than that of standard ones (see Fig. \ref{fig7}) and the oscillations occur inside the coarse grained cell of the width $\Delta=10$. 

To conclude, we see that, as long as the internal structure of the composite system is not addressed, collective effects in Bloch oscillations require interaction between the constituents. This observation matches the one from the previous MZI example. We discuss this problem in more details in the next section.


\section{Discussion}

We showed that the entangled states (\ref{thestate}) and (\ref{gauss2}) of two non-interacting particles can exhibit some composite features. However, in general the interaction is necessary to manifest the true compositeness. Here we discuss the above results and provide an explanation of why the fully composite behaviour cannot be manifested in non-interacting systems.

\subsection{No-signalling}

There is a simple rule, which is a variant of the no-signalling condition, that allows us to determine when an interaction is necessary to exhibit a composite behaviour. We start by noting that in case of non-interacting particles the evolution of the system can be written as a product of evolutions of its subsystems $U=e^{-i (H_a\otimes\openone+\openone\otimes H_b) t}=e^{-i H_a t} \otimes e^{-i H_b t} = U_a \otimes U_b$, where $H_a$ and $H_b$ are the Hamiltonians of particle $a$ and $b$, respectively. Even if the two particles are entangled, the individual evolution of each particle does not depend on the presence of the other particle, i.e., $\rho_a(t) = U_a \rho_a(0) U_a^{\dagger}$, where $\rho_a(0) = \text{tr}_{b}\{\rho_{ab}(0)\}$ and $\rho_{ab}(0)$ is the initial state of both particles (similar for particle $b$). Therefore, by looking at an individual particle, say particle $a$, one cannot say whether the particle $b$ is present or not. This is a form on no-signalling, since any ability to detect the presence of particle $b$ would lead to communication from $b$ to $a$ and, as far as we know, such communication is not possible if there is no interaction. 

Let us discuss the above in the context of two scenarios that we considered in the previous section. First, we focus on the MZI and ask what would happen if an entangled pair was capable of producing an interference pattern corresponding to the collective de Broglie wavelength. Moreover, we assume that this would be possible under our assumption that both constituents stay together. In this case the probability at the detector would be given by $\frac{1}{2}(1 + \cos 2\varphi)$. Next, imagine that due to some reason we are only able to detect the particle $a$. The particle $b$ travels together with $a$, but the detector does not register the presence of particle $b$. Nevertheless, the detector still reveals the interference pattern $\frac{1}{2}(1 + \cos 2\varphi)$.

However, imagine that just before the entry to the MZI somebody removed the particle $b$. Now, there is only a single particle in the MZI, therefore the interference pattern needs to correspond to a single-partite de Broglie wavelength. The probability at the detector becomes $\frac{1}{2}(1 + \cos \varphi)$. In particular, let us assume that $\varphi$ is fixed and is equal to $\pi$. In this case we would observe two fundamentally different situations, depending on whether $b$ is in the MZI, or not. If $b$ was present, then the particle $a$ would be registered at the detector with probability $1$, whereas if $b$ was absent then the particle $a$ would not be registered at this detector. This is a direct form of signalling, which is impossible as long as $a$ and $b$ do not interact.  

Similar signalling would occur in Bloch oscillations if oscillations with fractional periods were possible without interaction. If we prepared the system at the origin and observed that the particle $a$ returned to it after time $T_{BO}/2$, we would know that $b$ was around. However, if $a$ returned to the origin after time $T_{BO}$, then we would know that $b$ was not a part of the system. Therefore, by choosing whether to insert $b$ or not, one would be able to send signals to $a$, which is not possible without interaction. 

Therefore, we conclude that in any phenomena involving composite particles interaction is necessary whenever the phenomenon allows in principle for some form of signalling from one constituent to the other. Finally, we remark that such interaction can be simulated by post-selective measurements, i.e., by choosing to observe only certain outcomes. 

\subsection{Interaction vs Entanglement}

Next, let us provide another argument for the need of interaction and let us discuss the role of entanglement in case the interaction is not present. We are going to show that entanglement is a kind of resource that is needed to sustain the composite behaviour. 

In our previous work \cite{e23}, we argued that the evolution of a composite quantum particle from a localized state to a delocalized one requires production of entanglement. More precisely, consider a transformation of the form $c_{x_0}^{\dagger} \rightarrow \sum_x \alpha_x c_{x}^{\dagger}$, where $c^{\dagger}_x$ is a particle creation operator at position $x$ and $\sum_x |\alpha_x|^2=1$. If $c_x^{\dagger}=a_x^{\dagger}b_x^{\dagger}$ creates both particles $a$ and $b$ at position $x$, then the above transformation creates entanglement. This is because the initial state $c_{x_0}^{\dagger}|0\rangle$ is a product of two pure single-particle states $a_{x_0}^{\dagger}b_{x_0}^{\dagger}|0\rangle$. On the other hand, the final state is entangled, since individual particles are in a mixed state, e.g., $\rho_a = \sum_x |\alpha_x|^2 a^{\dagger}_x|0\rangle\langle 0|a_x$. Therefore, the above transformation is not possible without some interaction, which is required to generate entanglement.

The above observation might seem contradictory to the results obtained in the previous sections, where we showed that it is possible that two non-interacting particles can get delocalized and remain close to each other. However, the actual transformation behind this behaviour was of the form $c_{x_0}^{\dagger} \rightarrow \sum_x \alpha_x d_{x}^{\dagger}$, where 
\begin{equation}
c_{x_0}^{\dagger}= \sum_y \beta_y a_{x_0,y}^{\dagger}b_{x_0,y}^{\dagger},~~d_{x}^{\dagger}= \sum_y \gamma_{x,y} a_{x,y}^{\dagger}b_{x,y}^{\dagger}
\end{equation}
are operators that already imply some entanglement between $a$ and $b$. During such transformation the internal state of the composite particle, denoted by the index $y$, changes. The idea is that there are two types of entanglement in the system, the internal and the spatial one. While the entanglement encoded in the internal structure decreases, the spatial entanglement increases, so that the total entanglement stays the same. This means that the initial state of the component $a$ given by 
\begin{equation}
\rho_a(0)=\sum_y |\beta_y|^2 a_{x_0,y}^{\dagger}|0\rangle\langle 0| a_{x_0,y}
\end{equation}
needs to have the same von Neumann entropy $S(\rho)=\text{tr}\{\rho \log \rho\}$ as the corresponding final state 
\begin{equation}
\rho_a(t)=\sum_{x,y} |\alpha_x|^2|\gamma_{x,y}|^2 a_{x,y}^{\dagger}|0\rangle\langle 0| a_{x,y}.
\end{equation}

The above arguments can be interpreted as follows. The internal entanglement stored inside the composite particle is consumed in order to allow the system to get delocalized without falling apart, i.e., allow to preserve the close distance between the two components. The most critical transformation, after which decay is inevitable, is when the final operators are unentangled, i.e., $d_x^{\dagger}=a_x^{\dagger}b_x^{\dagger}$. In this case the whole internal entanglement is transformed into spatial entanglement. This interpretation is in accordance with the formula (\ref{entspread}), since it explains why the more initial entanglement between the particles, the farther the composite particle can spread. 

\subsection{Composite momentum on a lattice}

Here we discuss yet another important aspect of the interaction that is particularly important for Bloch oscillations with fractional periods. For a particle moving on a lattice, with a lattice constant set to one,  we get $k\in [-\pi,\pi]$, i.e., momentum is confined to the first Brillouin zone. This leads to the Bloch theorem, which implies $\psi(k)\equiv \psi(k+n2\pi)$, i.e., the state does not change if the momentum is shifted by a multiple of $\hbar 2\pi$. 

Now, consider two particles moving on a lattice. Their corresponding momenta, $\hbar k_1$ and $\hbar k_2$, are both confined to the first Brillouin zone. The centre of mass of both particles is described by $K=k_1 + k_2$. If the centre of mass were to describe a single composite particle, then $K$ should be also confined to the first Brillouin zone. Let us consider a simple example. If $k_1=k_2=\pi$, then $K=2\pi$. On the other hand, if $k_1=k_2=0$, then $K=0$. If $K$ were in the first Brillouin zone, then the above cases would be indistinguishable and would result in the same state. However, in general these two cases lead to two different states. Therefore, we need to find a mechanism which makes these states indistinguishable. We are going to show that such a mechanism is provided by the interaction.

We consider a finite one-dimensional lattice with $d$ sites (without loosing generality we assume $d$ is even), lattice constant set to one, and periodic boundary conditions. First, we represent the interaction Hamiltonian (\ref{int}) in the momentum basis
\begin{equation}
a_x^{\dagger}=\frac{1}{\sqrt{d}}\sum_{k=1}^d e^{ikx}\tilde{a}_k^{\dagger},
\end{equation}
where $\tilde{a}_k^{\dagger}$ creates a particle with momentum $\hbar k$. The momentum has $d$ different values and $k=-\pi +\frac{2\pi}{d}, -\pi + \frac{4\pi}{d},-\pi + \frac{6\pi}{d},\ldots,\pi$. This leads to
\begin{equation}\label{int2}
H_{int}=\frac{\gamma}{d} \sum_{k,l,q} \tilde{a}_{k+q}^{\dagger}\tilde{b}_{l-q}^{\dagger}\tilde{a}_{k}\tilde{b}_{l}.
\end{equation}
The above operator acts on two particles with momenta $\hbar k$ and $\hbar l$, respectively, and creates a superposition
\begin{equation}\label{stateK}
|\psi_{K=k+l}\rangle = \frac{1}{\sqrt{d}}\sum_q \tilde{a}_{k+q}^{\dagger}\tilde{b}_{l-q}^{\dagger} |0\rangle.
\end{equation}
Therefore, it causes the exchange of momentum between the particles, but the total momentum is conserved. Note, that $|\psi_{K=k+l}\rangle$ are degenerate eigenstates of $H_{int}$ with the eigenvalue $\gamma$. These eigenstates have a well defined momentum of the centre of mass.  
 
Now, consider the action of $H_{int}$ on the states from the above example. We get
\begin{eqnarray} 
H_{int}\tilde{a}_{\pi}^{\dagger}\tilde{b}_{\pi}^{\dagger} |0\rangle &=& \frac{\gamma}{d} \left( \tilde{a}_{\pi}^{\dagger}\tilde{b}_{\pi}^{\dagger} + \tilde{a}_{\pi+\frac{2\pi}{d}}^{\dagger}\tilde{b}_{\pi-\frac{2\pi}{d}}^{\dagger} + \ldots \right)|0\rangle, \\
H_{int}\tilde{a}_{0}^{\dagger}\tilde{b}_{0}^{\dagger} |0\rangle &=& \frac{\gamma}{d} \left( \tilde{a}_{0}^{\dagger}\tilde{b}_{0}^{\dagger} + \tilde{a}_{\frac{2\pi}{d}}^{\dagger}\tilde{b}_{-\frac{2\pi}{d}}^{\dagger} + \ldots \right)|0\rangle.
\end{eqnarray}
However, $\tilde{a}_{\pi+n\frac{2\pi}{d}}^{\dagger} \equiv \tilde{a}_{-\pi+n\frac{2\pi}{d}}^{\dagger}$ because the momentum of individual particles is confined to the first Brillouin zone. Therefore $\tilde{a}_{\pi+\pi}^{\dagger}\tilde{b}_{\pi-\pi}^{\dagger} \equiv \tilde{a}_{0}^{\dagger}\tilde{b}_{0}^{\dagger}$ and $\tilde{a}_{0+\pi}^{\dagger}\tilde{b}_{0-\pi}^{\dagger} \equiv \tilde{a}_{\pi}^{\dagger}\tilde{b}_{\pi}^{\dagger}$, i.e., the exchange of momentum between the particles can swap the state $\tilde{a}_{\pi}^{\dagger}\tilde{b}_{\pi}^{\dagger} |0\rangle$ into $\tilde{a}_{0}^{\dagger}\tilde{b}_{0}^{\dagger} |0\rangle$ and vice versa. Moreover, 
\begin{equation}
H_{int}\tilde{a}_{\pi}^{\dagger}\tilde{b}_{\pi}^{\dagger} |0\rangle=H_{int}\tilde{a}_{0}^{\dagger}\tilde{b}_{0}^{\dagger} |0\rangle = \frac{\gamma}{\sqrt{d}}|\psi_{K=2\pi}\rangle = \frac{\gamma}{\sqrt{d}} |\psi_{K=0}\rangle.
\end{equation}
As a result, in case of interaction momentum of the centre of mass is also confined to the first Brillouin zone, which is a prerequisite for a composite particle behaviour on a lattice.

Finally, let us consider the action of the operator $e^{-itV}$, where $V$ is given by (\ref{BOH}), on $|\psi_{K=k+l}\rangle$. We get $e^{-itV}|\psi_{K=k+l}\rangle = |\psi_{K=k+l-2t\eta}\rangle$, i.e., the value $K$ is shifted by $-2\eta t$. However, since in case of interaction $K$ is confined to the first Brillouin zone, we observe Bloch oscillations with a fractional period $T_{BO}=\frac{\pi}{\eta}$.

\subsection{Final remarks}

Let us summarize the general features of composite particle-like dynamics of non-interacting entangled pairs. Firstly, we are not able to observe these types of dynamics for which the form of the reduced density matrix of one subsystem depends on the presence of the other subsystem. In addition, in order to fulfil the requirement that the constituents are close to each other, we allow evolutions which only minimally affect the distance between the two particles. This, however, can be achieved by a proper state preparation, like (\ref{thestate}) or (\ref{gauss2}), where the relative distance is described by a Gaussian state with an acceptably large variance (corresponding to the resolution of the measurement apparatus). For example, one can consider states of the form $\psi(x_1,x_2)=\varphi(x_1+x_2)e^{-\frac{(x_1-x_2)^2}{4\sigma^2}}$, where $\varphi(x_1+x_2)$ is an arbitrary wave-function. In this case, for times $t<\frac{m}{\hbar}\sigma^2$, the time evolution can be approximated as $\psi(x_1,x_2,t) \approx \varphi(x_1+x_2,t)e^{-\frac{(x_1-x_2)^2}{4\sigma^2}}$. This implies that the wave-function is approximately non-vanishing for $x_2 \in (x_1 - \sigma,x_1 + \sigma)$. If the precision of the detectors is of the order $\sigma$, or worse, then effectively $x_2 \approx x_1$ and the system is described by a position of a single particle. But, because the presence of the other particle is undetectable, the above implies that the composite system behaves as a single elementary particle, not as a composite particle. This was confirmed by the MZI and the Bloch oscillation numerical simulations. 


{\it Acknowledgements.} This work was supported by the National Science Centre in Poland through the NCN Grant No. 2014/14/E/ST2/00585. In addition, SYL is supported by Basic Science Research Program through the National Research Foundation of Korea (NRF) funded by the Ministry of Education (2018R1D1A1B07048633).

\section*{}

\end{document}